\documentclass[a4]{aa}
\usepackage{natbib}
\usepackage{graphicx}
\newcommand{\logg}{\mbox{$\log g~$}}

\begin{document}
\title{Search for associations containing young stars
\thanks {Based on observations collected at the ESO - La Silla and 
   at the LNA-OPD}}
\subtitle{III. Ages and Li abundances}
\author{L. da Silva \inst{1}, C. A. O. Torres\inst{2}, R. de la
  Reza\inst{1}, G. R. Quast\inst{2}, C.  H. F. Melo\inst{3} and M. F. Sterzik\inst{3}} 
\offprints{licio@on.br}
\institute{Observat\'orio Nacional-MCT, Rio de Janeiro, Brazil
\and
Laborat\'orio Nacional de Astrof\'{i}sica-MCT, Itajub\'a, Brazil
\and
European Southern Observatory, Alonso de Cordova 3107, Casilla 19, Santiago, Chile}
\date{received ...; accepted...}
\abstract{Our study is a follow-up of the SACY project, an extended high spectral resolution survey of
more than two thousand optical counterparts of X-ray sources in the Southern Hemisphere
targeted to search for young nearby associations. Nine associations have either been newly identified,  
or had their member list better defined. Groups belonging to the
Sco-Cen-Oph complex are not considered in the present study. }
{These nine associations, with ages
{  between about 6\,Myr and 70\,Myr}, form an excellent sample
to study the Li depletion in the pre-main sequence(PMS) evolution.  
In the present paper we investigate the use of Li abundances  as an
independent clock to constrain the PMS evolution.}
{Using our  measurements of the equivalent widths of the Li resonance line 
and assuming  fixed metallicities and microturbulence,
we have calculated the LTE Li abundances for 376 members of different young associations.
In addition we considered the effects of their projected stellar rotation.}
{We present the Li depletion as function of age in the first hundred million years for the first time 
for the most extended sample of Li abundances in young stellar associations. }
{A clear  Li depletion can be measured in the temperature range from
   5000\,K to 3500\,K for the age span covered by the nine associations studied in this paper. 
   The age sequence based on the
Li-clock agrees well with the isochronal ages, $\epsilon$Cha association being the only 
possible exception.
The lithium depletion patterns for the associations presented here resemble those
of the young open clusters with similar ages, strengthening  the notion that the members proposed for these loose 
young associations have
indeed a common physical origin.
The observed scatter in the  Li abundances
   hampers the use of Li  to determine reliable ages for individual stars. 
   For velocities above 20\,km\,s$^{-1}$ rotation seems to play an important role inhibiting the  Li depletion. 
}
\authorrunning{L. da Silva  et al.}
\titlerunning{Age and Li in young associations}
\maketitle
%
%
\section{Introduction} 
In \cite{torres06} (hereafter Paper\,I) we report the
results of a high-resolution optical spectroscopic survey aimed at searching
for associations containing young stars (SACY) among optical
counterparts of ROSAT All-Sky X-ray sources in the Southern Hemisphere.
There we present  the catalog resulting from the survey. 
We describe the convergence method developed to search for members of an association
and a corresponding membership probability model.
A membership to an association is defined  in  the hexa-dimensional space formed
by the (UVW) velocity space and the (XYZ) spatial coordinates distribution.
We take also into account the position in the HR diagram, 
eliminating very discrepant stars. 
Finally we check each member proposed confronting  its Li content with the
Li distribution of the association.  
The $\beta$~Pictoris Association ($\beta$PA) is presented as an example of the method outlined in Paper\,I.

In  paper\,I we also present
the Li abundance analysis of the $\beta$PA, in order
to confirm its youth.  
In contrast to open clusters where Li abundances have been studied over more 
than one decade \citep[see][]{pallavicini00},
the results of Paper\,I was the first analysis of this kind for a young association.

Using the method described in Paper\,I, \cite{torres08} (hereafter Paper\,II)
defined nine new young associations, namely,
 $\epsilon$~Chamaleontis ($\epsilon$ChA), TW~Hydrae (TWA),
$\beta$~Pictoris, Octants (OctA), Tucana-Horologium (THA),
 Columba (ColA), Carina (CarA),
Argus (ArgA),  and AB~Doradus (ABDA). 
The present work continues along the same lines
and aims at deriving the distribution
of Li abundances for each of the nine associations
resulting from a consolidated list of members.
As these associations are
young, covering ages from about 5\,Myr up
to that of the Pleiades, they form an interesting "laboratory"
to study the Li depletion with age,  as done for some open clusters
\citep{randich01,jeffries06}.

\section{Sample}

In Table\,1 we present some properties of the young associations studied in this paper derived in  Paper\,II totaling 376 stars for which Li abundances were measured.
Although the data are mainly from Paper\,I,
a few additional data obtained meanwhile are included and will be published in forthcoming papers.

These new observations allowed us to refine the definitions for some of the associations.
For the ColA,  we are able to obtain
a  similar but more consistent solution with a few changes in the  member list  with respect  to those of
Paper\,II   (three stars are now rejected and six new ones are included). 

For the ABDA, three new members have been proposed, one of them, HD\,82879, previously proposed to the  $\epsilon$ChA. 
HD\,53842, proposed by \cite{zuckermansong04} as a member of the THA, was
previously rejected in Paper\,II  due to a compilation error.
Its re-introduction now as a member proposed for the THA has no other consequences for
the mean values of this association\footnote{With these updates of the THA and the ColA,
only 11 of the 50 stars listed by \cite{zuckermansong04} are not found by us
as high probability members of one of the GAYA associations \citep[see section 3 of][]{torres08}.}.

As explained in Paper\,II,  IC~2391 members were incorporated in the ArgA member list.
Similarly
 members of the  open cluster $\eta$~Cha have been put together with
 the $\epsilon$ChA members forming an unique group. 
 The link between young loose associations and some open clusters is 
 going to be discussed in forthcoming papers.

 
\begin{table}
\caption{Number of members (N), number of Li eliminated stars ("intruders", n), average distance (in parsecs) and
age (in Myr) of the considered associations}
\label{table:assoc}
\begin{center} 
\begin{tabular}{||lrrrr||}
\hline\hline
Association & N & n  & distance & Age\\
\hline
$\epsilon$~Chamaleontis ($\epsilon$ChA)&23 &0& 93 -- 123  &\,\,\,6\\
TW~Hydrae (TWA)     & 22 &0& 28 -- \,\,\,73  & \,\,\,8\\  
$\beta$~Pictoris ($\beta$PA) &48&2& 10 -- \,\,\,80&10\\
Octans (OctA)         & 15 &0& 82 -- 175 &20?\\
Tucana-Horologium (THA)     &45 &1& 37 -- \,\,\,78   &30\\
Columba (ColA)        &44 &0& 35 -- 150 &30\\
Carina (CarA)        & 23 &1& 45 -- 160 &30\\
Argus (ArgA)       & 64 &1& 29 -- 164&40\\  
AB~Doradus (ABDA)     & 92 &4& 7 -- 143 &70\\
\hline\hline    
\end{tabular}  
\end{center}
\end{table}

\section{Li Abundance determinations}
The observations were made using the
FEROS spectrograph \citep{kaufer99} at la Silla, ESO, and the Coud\'e
spectrograph of the  1.62\,m telescope of the Observat\'orio do Pico dos Dias,
LNA, Brazil (see Paper\,I for details).
The Li abundances (A$_{\rm Li}$) of the stars, in dex, in the system log\,A(H)\,=\,12,
where A(H) is the H abundance (A$_{\rm Li}$\,=\,log{N(Li)/N(H)}\,+\,12),  
were determined using the programs of M. Spite, of the
Paris-Meudon Observatory.  Our method is similar to that used for the $\beta$PA
in paper\,I. The main difference is that we now
apply
atmospheric models of Kurucz and Castelli
(www.user.oat.ts.astro.it/castelli) instead of  those
of \cite{gustafsson75} used in paper\,I.

The  A$_{\rm Li}$ were determined from the resonance line at $\lambda\,6708$.
The method consists in calculating the theoretical equivalent widths of the Li
line  (EW$_{\rm Li}$) and comparing them with the  corresponding observed ones.  
The A$_{\rm Li}$ is changed until the difference between the
calculated and the observed EW$_{\rm Li}$ is smaller than 0.2\,m{\AA}. 
The line was considered to be formed only by the  $^7$Li isotope. In the computation
of the synthetic profile we take into account the four components of the $^7Li$
resonance line, being the wavelengths and the oscillators strengths given by
Andersen et al. (1984). (Wavelengths: $\lambda\,6707.754, \lambda\,6707.766,
\lambda\,6707.904, \lambda\,6707.917$; and log\,$gf$: $-0.430, -0.209, -0.733,
-0.510$, respectively.) 
   
Effective temperatures were obtained from the
photometric and spectroscopic data available. The calibrations used were mainly those of
\cite{kenyon95} and \cite{schmidt-kaler82}.
Some additional information was included from \cite{bessel79}
and from \cite{ducati01}.
If a reliable Cousins  (V-I)$_c$ color index was available, either from
our observations, from Hipparcos or from other sources in the literature,
this was used to derive T$_{\rm eff}$. In the absence of  (V-I)$_c$, we used the
Johnson (B-V), mainly derived from TYCHO-2 but also obtained from various
sources in the literature. We considered the (B-V) colors from TYCHO-2  reliable only
for stars  brighter than magnitude 10. Finally, if no reliable photometry
was available, we used the spectral type to obtain the effective
temperature.

%

The other model parameters were kept fixed:
metallicity at [Fe/H]\,=\,0.1 \citep[see][]{castilho05};
gravity \logg was fixed at 4.5 for the dwarfs,
and  at 4.0 for the subgiants, according to the spectral
classification of Paper\,I.
The  microturbulence velocity was
fixed at 1.5\,km~s$^{-1}$ for all stars.

\subsection{Error Analysis}

In order to assess our internal errors we study the variations of the Li abundance 
as a function of model parameters and the equivalent width (Table~\ref{table:error}). 
Our main error source is the  effective temperature. The
unknown parameters, [Fe/H] and the microturbulence velocity are less
important, and the values used are good estimates. 
The small sensitivity of the Li abundance with the
microturbulence velocity may be
surprising, but it can be explained by the use of the fine structure of the Li
line and by the fact that each line component is a weak line, thus not making very sensitive contribution to
this parameter. 
An error of 10\% 
for the EW is perhaps optimistic for the
weak lines. An increase of the error of the  EW to 20\% 
results in a variation of 
A$_{\rm Li}$ of 0.08 at T$_{\rm eff}$\,=\,4000\,K and of 0.09 at 6100\,K.
From Table\,3 we can say that our internal errors are smaller than 0.2, sufficient to reach our goal. 
As  can be seen from our abundance results in 
the figures, even a difference as high as 0.2 does not modify any of our conclusions.

How does the choice of different models change
our results? In order to address  this question, we compare our results using  Kurucz models with those
using  the  Uppsala group models, which were used in paper\,I.
Using atmospheric models calculated with the MARCS code, developed by the
Uppsala group (http://marcs.astro.uu.se/) \citep[see][]{gustafsson75}, we get
A$_{\rm Li}$\,0.09 larger at  4000\,K than using Kurucz models.
At 6000\,K  the difference is 0.07, in the same direction.
For homogeneity purposes, we used Kurucz models in our analysis
because they begin at 3500\,K whereas MARCS models begin only at  4000\,K.
Nevertheless, this shows how sensitive  the use of different
A$_{\rm Li}$ from different authors can be.
In this case the difference between our A$_{\rm Li}$  results and those of other
authors could  even be larger than 0.2, which we adopt as internal
error. 

\begin{figure}
\begin{center}
\resizebox{0.8\hsize}{!}{\includegraphics{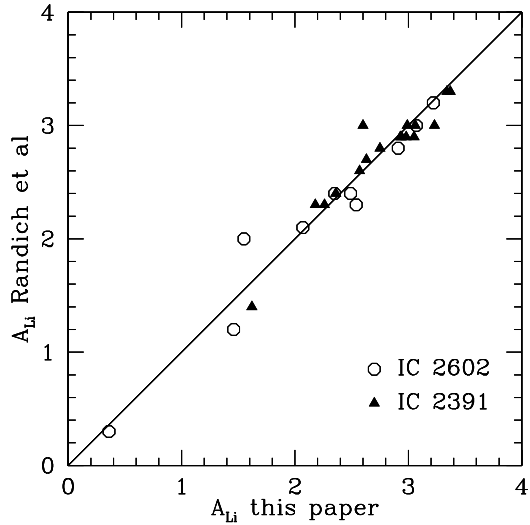}}
\end{center}
\caption{Comparison between our results and those of \cite{randich01} for the stars  in IC2391 and IC2602.}
\label{fig:licio_sofia}
\end{figure}

As an additional test,
we compared our abundances for the IC~2391 (considered as ArgA members) with those from
\cite{randich01}. In addition using the EW from these authors, we also computed the Li abundances for the IC2602 members as described above.
The agreement between  both Li abundances is very good (Figure~\ref{fig:licio_sofia})
validating our method.

\begin{table}
\caption{Variation of Li abundance as a function of EW and model parameters.
  $\Delta A(Li)_1$ is the variation at 4000\,K and $\Delta A(Li)_2$ at 6100\,K}
 \label{table:error}
\begin{center}
\begin{tabular}{||lrrr||}
\hline\hline
Parameter & change   & $\Delta A(Li)_1$ & $\Delta A(Li)_2$ \\
 \hline
T$_{\rm eff} [K]$ & $\pm$100 & $^{+0.18}_{-0.19}$ & $^{+0.08}_{-0.07}$ \\
\logg        & $\pm$0.5 & $^{+0.16}_{-0.11}$ & $^{+0.12}_{-0.06} $\\
microturbulence  [km s$^{-1}$] & $\pm$0.5 &$^{+0.05}_{-0.05}$ & $^{+0.02}_{-0.02}$ \\
EW            & $\pm$10\%   & $^{+0.04}_{-0.05}$ &$^{+0.05}_{-0.05}$ \\
$[Fe/H]$ & $\pm$0.1 & $^{-0.03}_{+0.04}$ & $^{-0.01}_{+0.02}$\\
\hline\hline    
\end{tabular}  
\end{center}
\end{table}


\section{Results and discussion}

\begin{figure*}
\begin{center}
\resizebox{0.32\hsize}{!}{\includegraphics[bb=-24 182 629 590,clip]{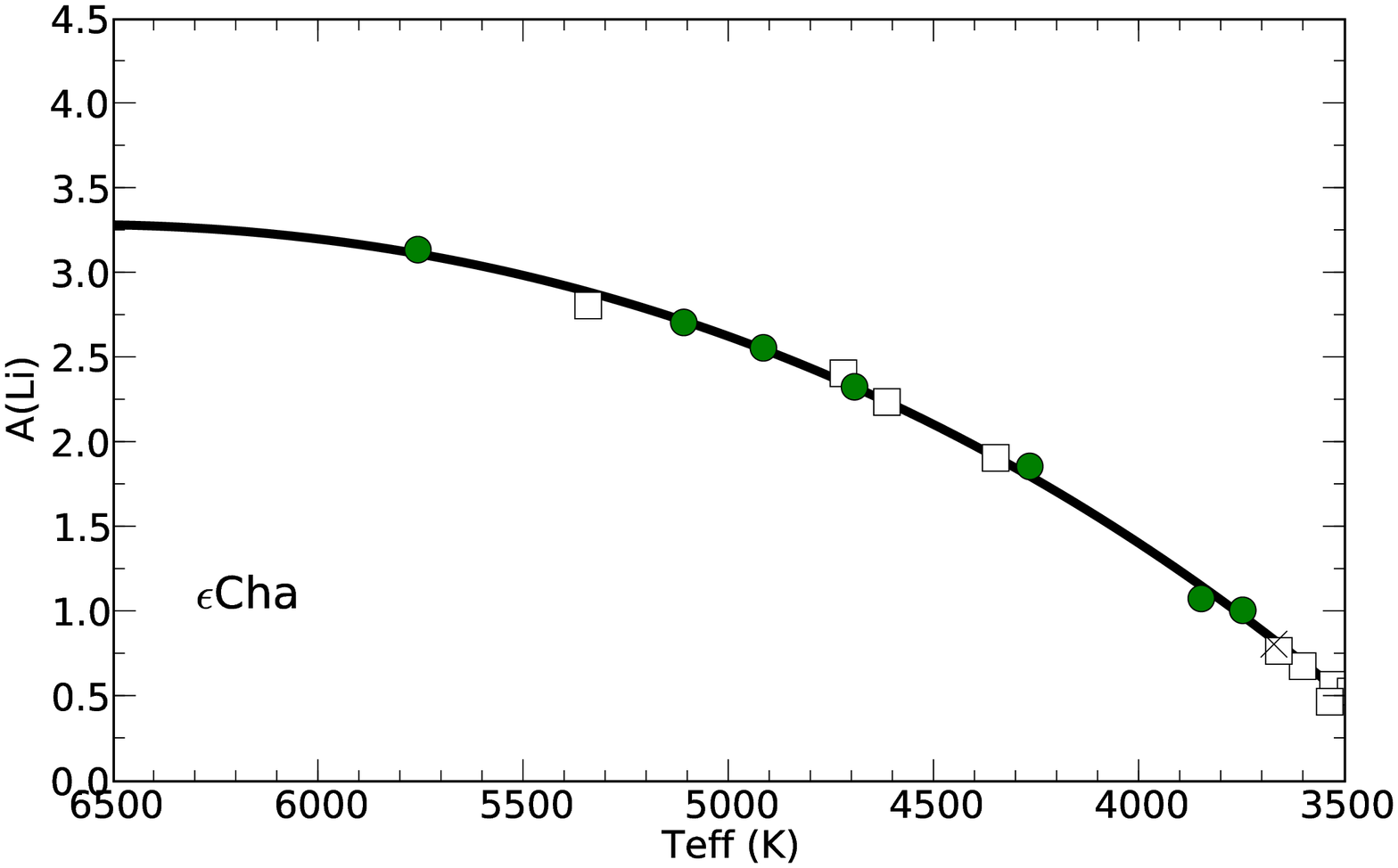}}
\resizebox{0.32\hsize}{!}{\includegraphics[bb=-24 182 629 590,clip]{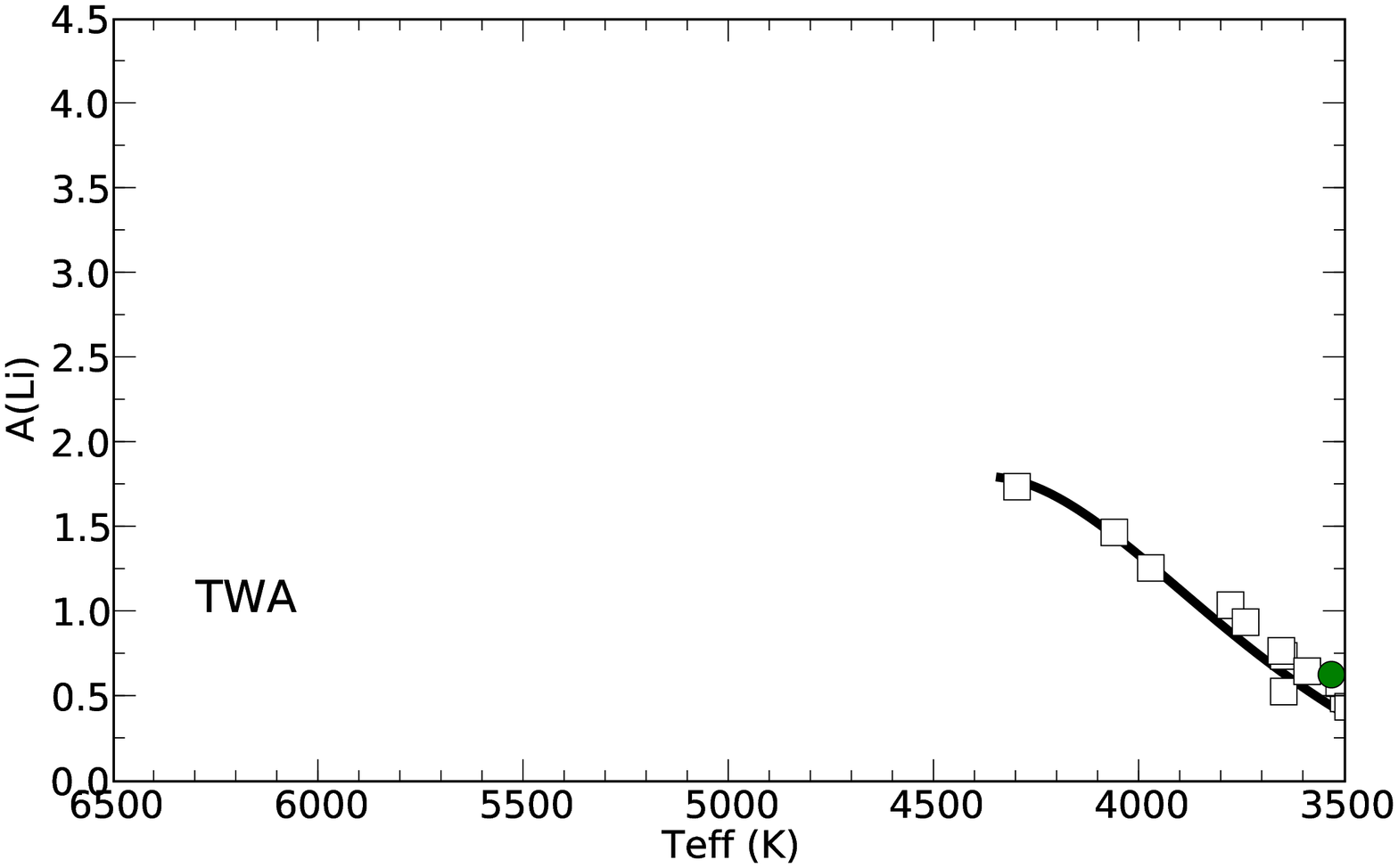}}
\resizebox{0.32\hsize}{!}{\includegraphics[bb=-24 182 629 590,clip]{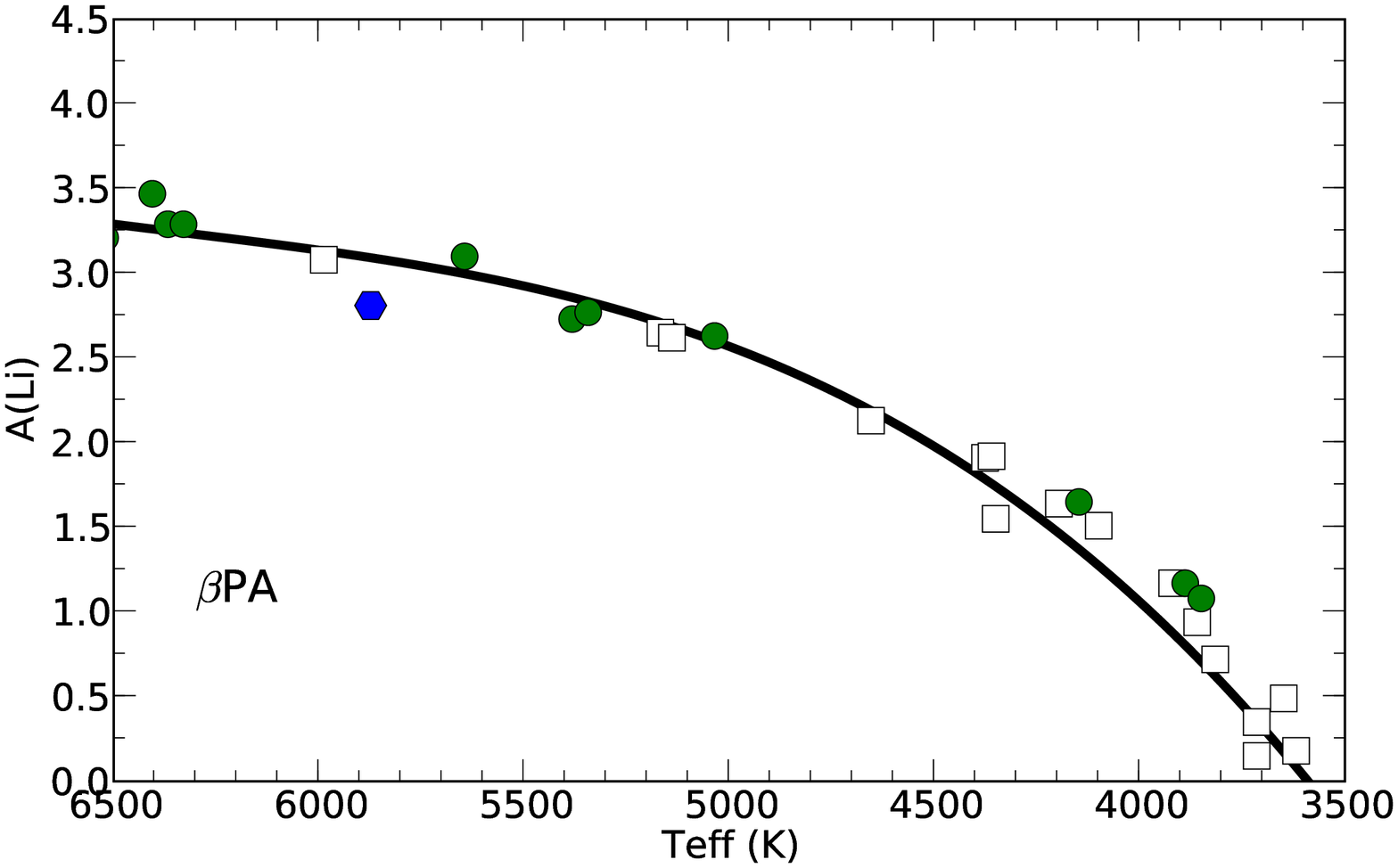}}
\resizebox{0.32\hsize}{!}{\includegraphics[bb=-24 182 629 590,clip]{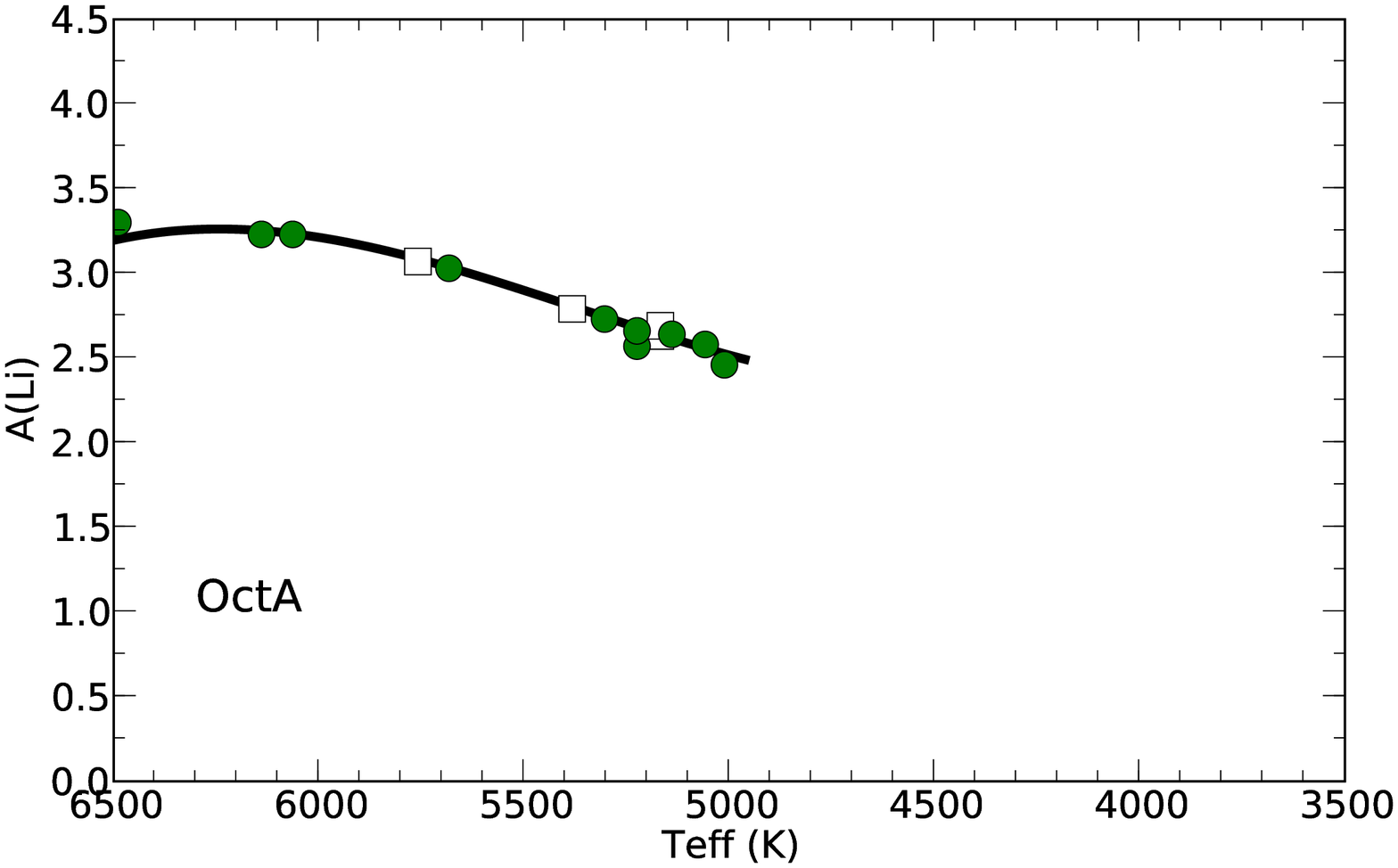}}
\resizebox{0.32\hsize}{!}{\includegraphics[bb=-24 182 629 590,clip]{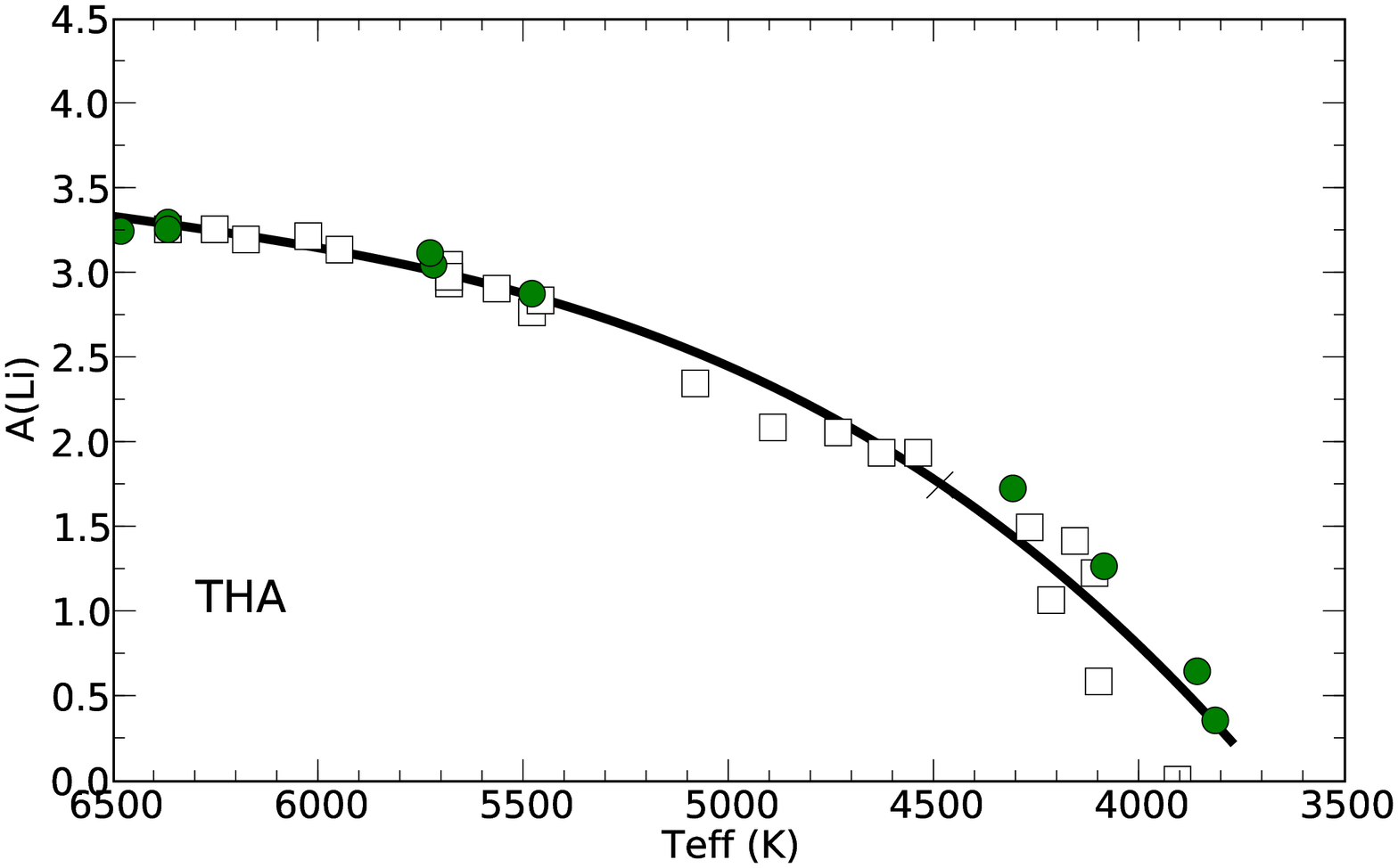}}
\resizebox{0.32\hsize}{!}{\includegraphics[bb=-24 182 629 590,clip]{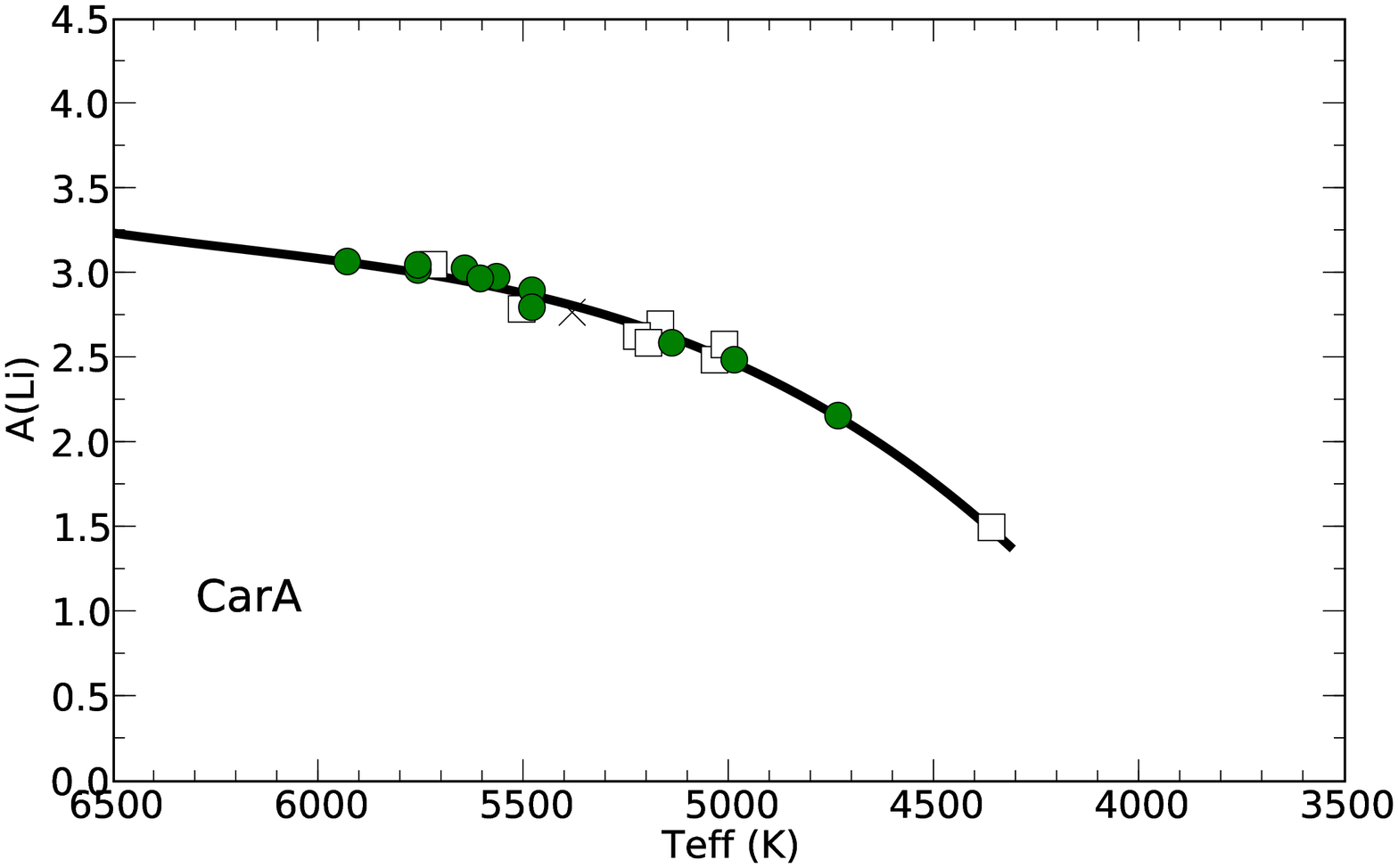}}
\resizebox{0.32\hsize}{!}{\includegraphics[bb=-24 182 629 590,clip]{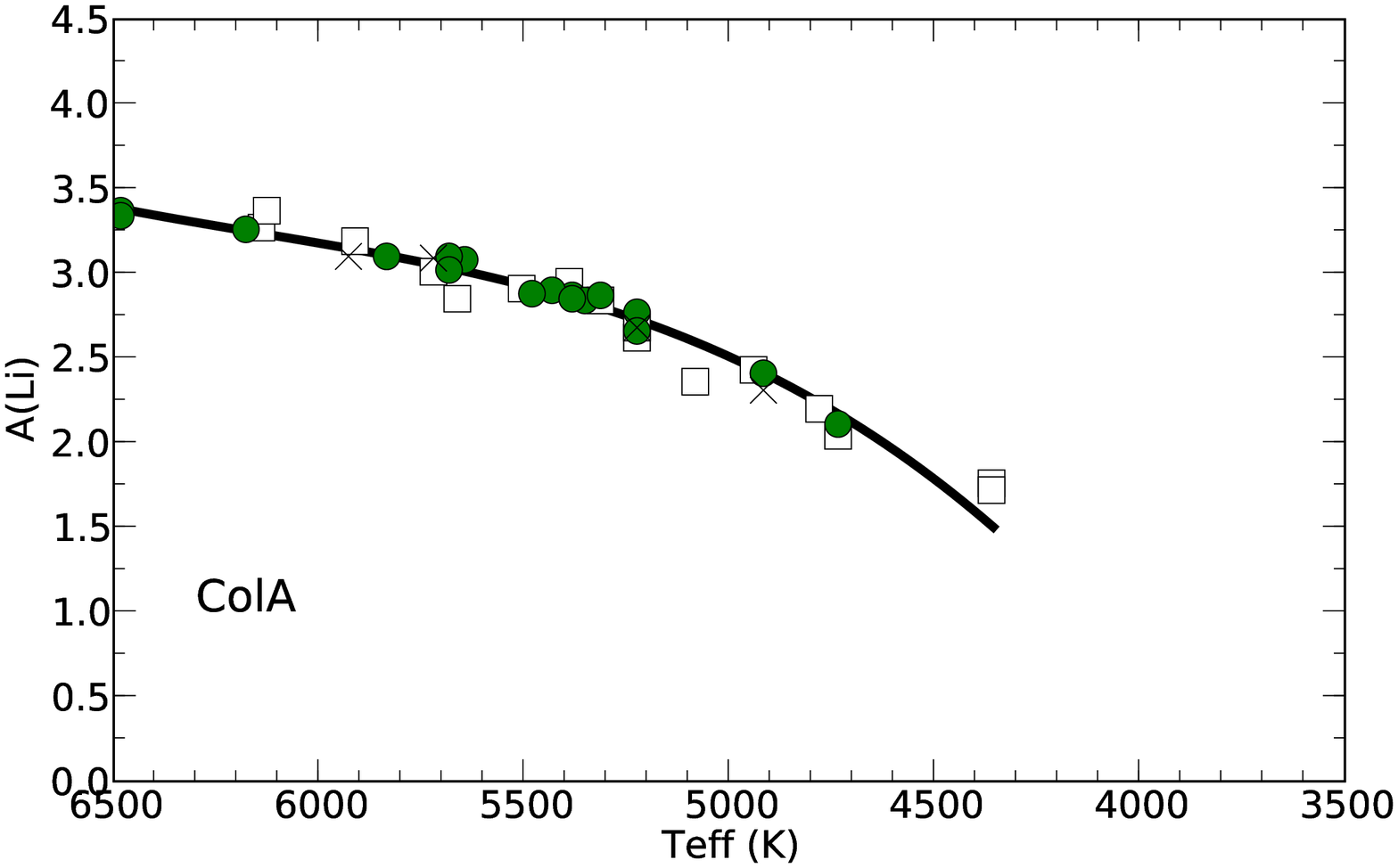}}
\resizebox{0.32\hsize}{!}{\includegraphics[bb=-24 182 629 590,clip]{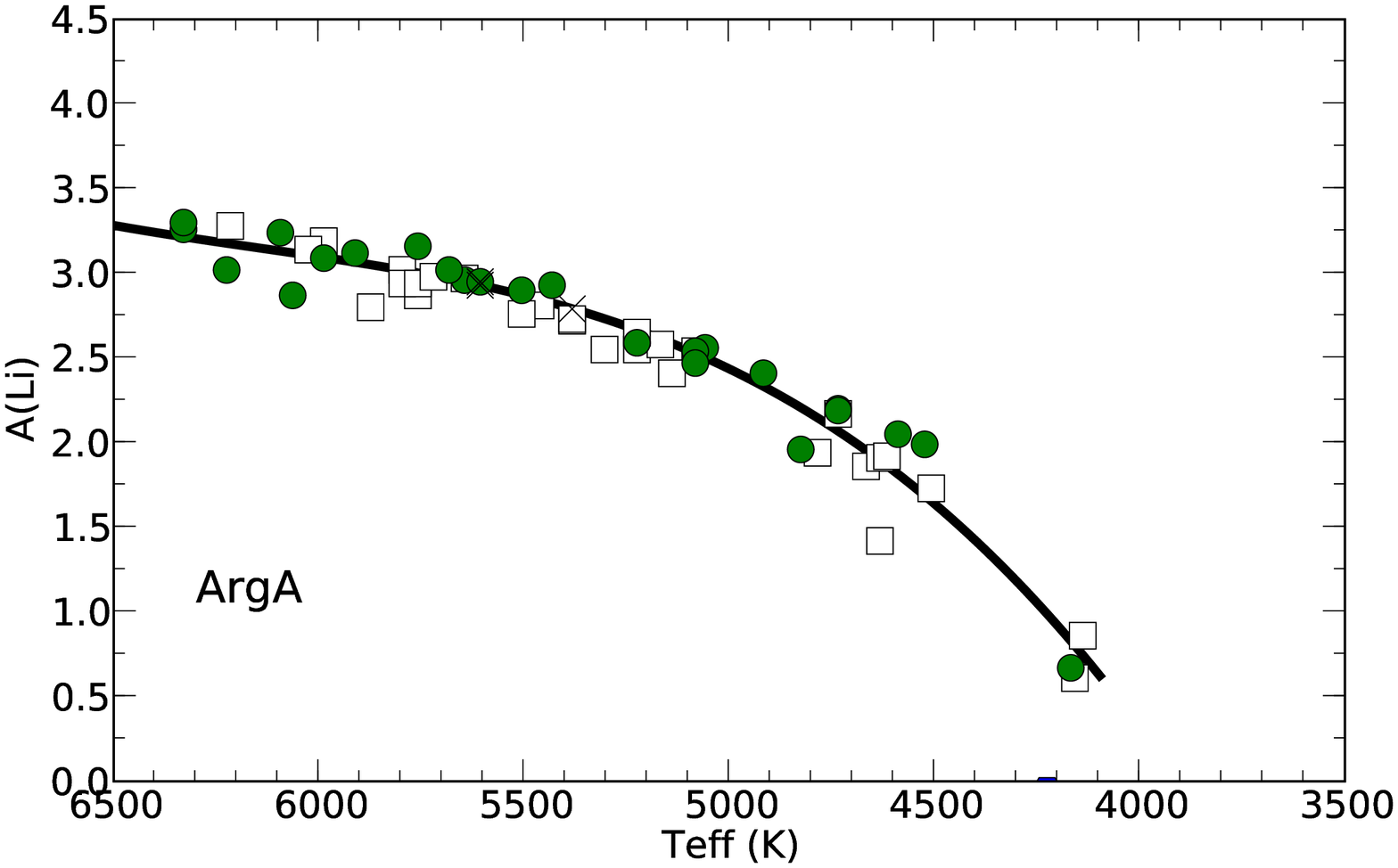}}
\resizebox{0.32\hsize}{!}{\includegraphics[bb=-24 182 629 590,clip]{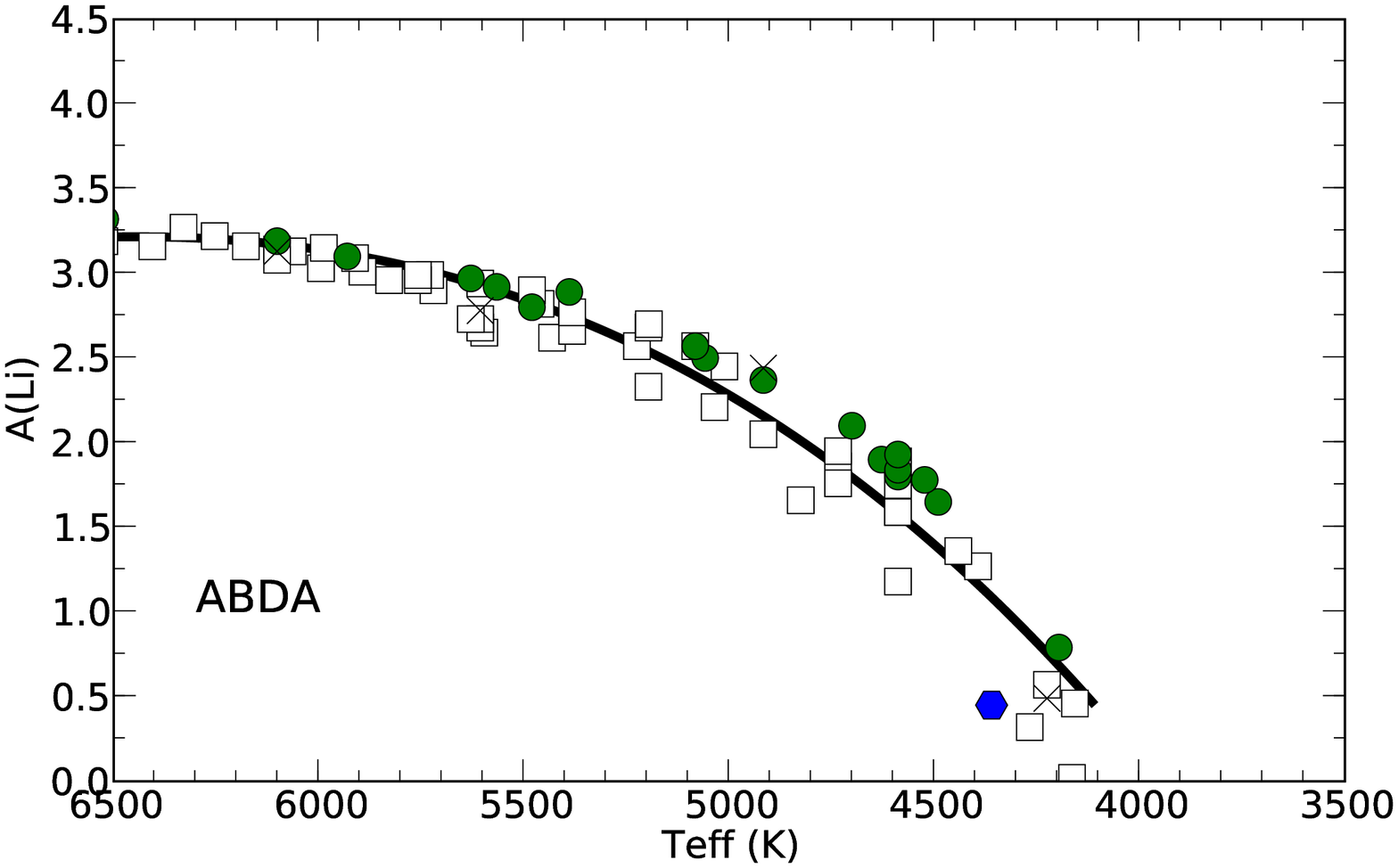}}

\end{center}
\caption{Lithium depletion pattern for all nine associations presented in this paper. Plots 
are arranged in the age sequence of Table~\ref{table:assoc}.
Stars rotating slower (faster)  than 20\,km\,s$^{-1}$ are shown as open squares (filled circles).
Stars whose $v\sin i$ could not
be determined are plotted as crosses. Filled hexagons are Li intruders.
A
4th-degree polynomial fit of the data is shown as a solid line.
}
\label{fig:ldp}
\end{figure*}

Table\,1 summarizes some properties given in Paper\,II for the nine  young associations. This table contains the proposed ages,  the most
important parameter for our Li evolution study.
Distance is a rather meaningless quantity for these nearby associations (due to their proximity
their members {  have a wide range of distances}).
The Li abundances determined for all high probability members, in all  376 stars, 
are given in  Tables\,4 to 12.  
The tables contain the identifications of the members of each association, 
their coordinates, the EW$_{\rm Li}$, the T$_{\rm eff}$, the A$_{\rm Li}$  
and  the projected stellar rotational velocities ($v\sin(i)$). 
More details about the association memberships can be found in  Paper\,II.

Stars cooler than 3500\,K are not covered by the Kurucz models and are given here for
 reference only. 
Those values are calculated with extrapolated models, and they  have errors 
potentially larger than those considered below.
Those objects have not been considered  in our figures and discussion. 

\subsection{Lithium Depletion Pattern and open cluster ages}
The Lithium Depletion Pattern (LDP) for all nine young associations studied in this paper is shown in Figure~\ref{fig:ldp}. 
For each association stars were divided into two groups according to their $v\sin(i)$. 
Stars rotating slower than 20\,km\,s$^{-1}$ are shown as open squares, whereas those 
rotating faster than 20\,km\,s$^{-1}$ are marked as filled circles. Stars whose $v\sin i$ could not
be determined are plotted as crosses.
Along with the derived abundances and effective temperatures, a
4th-degree polynomial fit of the data is shown as a solid line. This line defines the LDP for each association.





\begin{figure*}
\begin{center}
\resizebox{0.45\hsize}{!}{\includegraphics{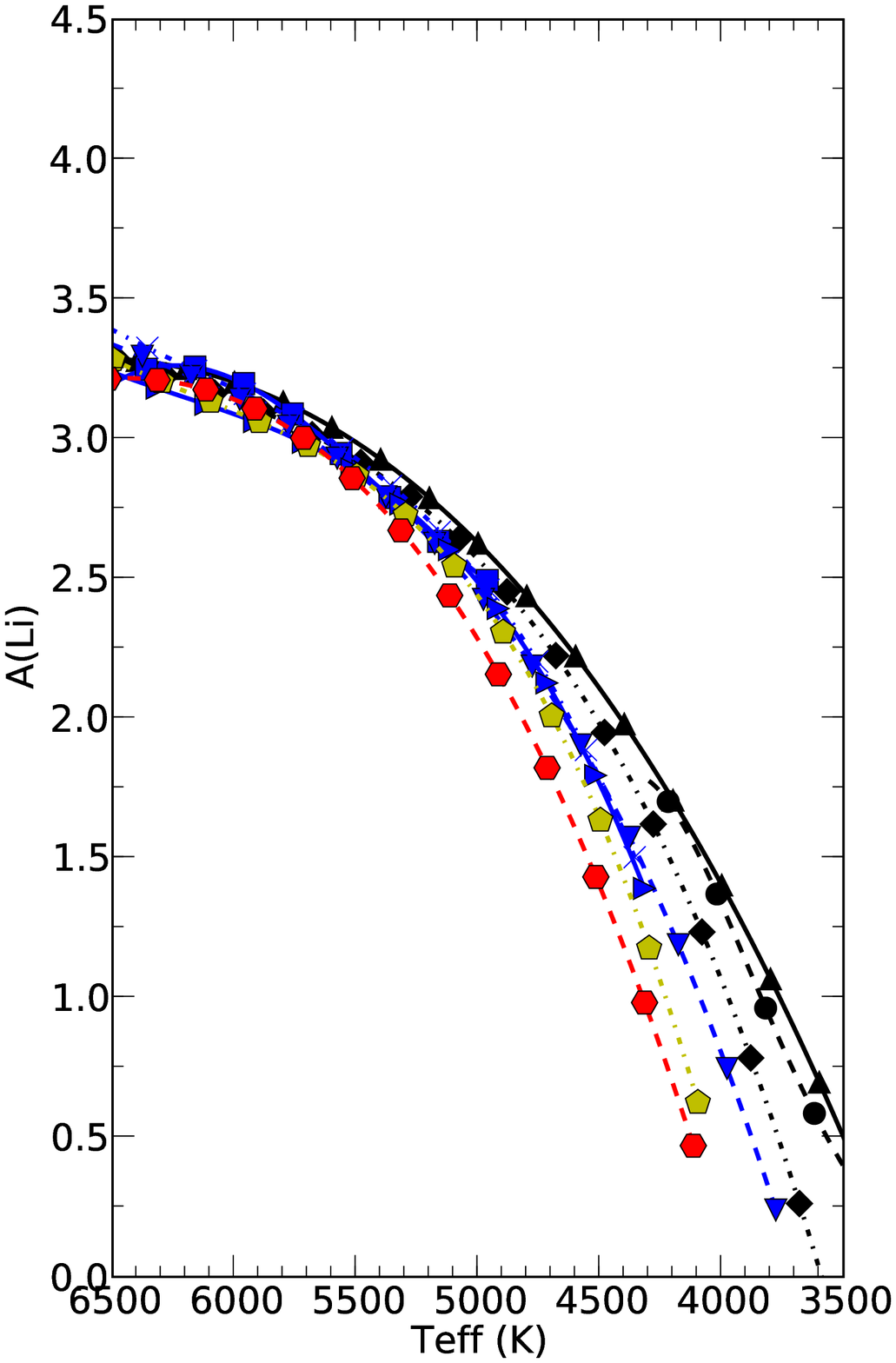}}
\resizebox{0.45\hsize}{!}{\includegraphics{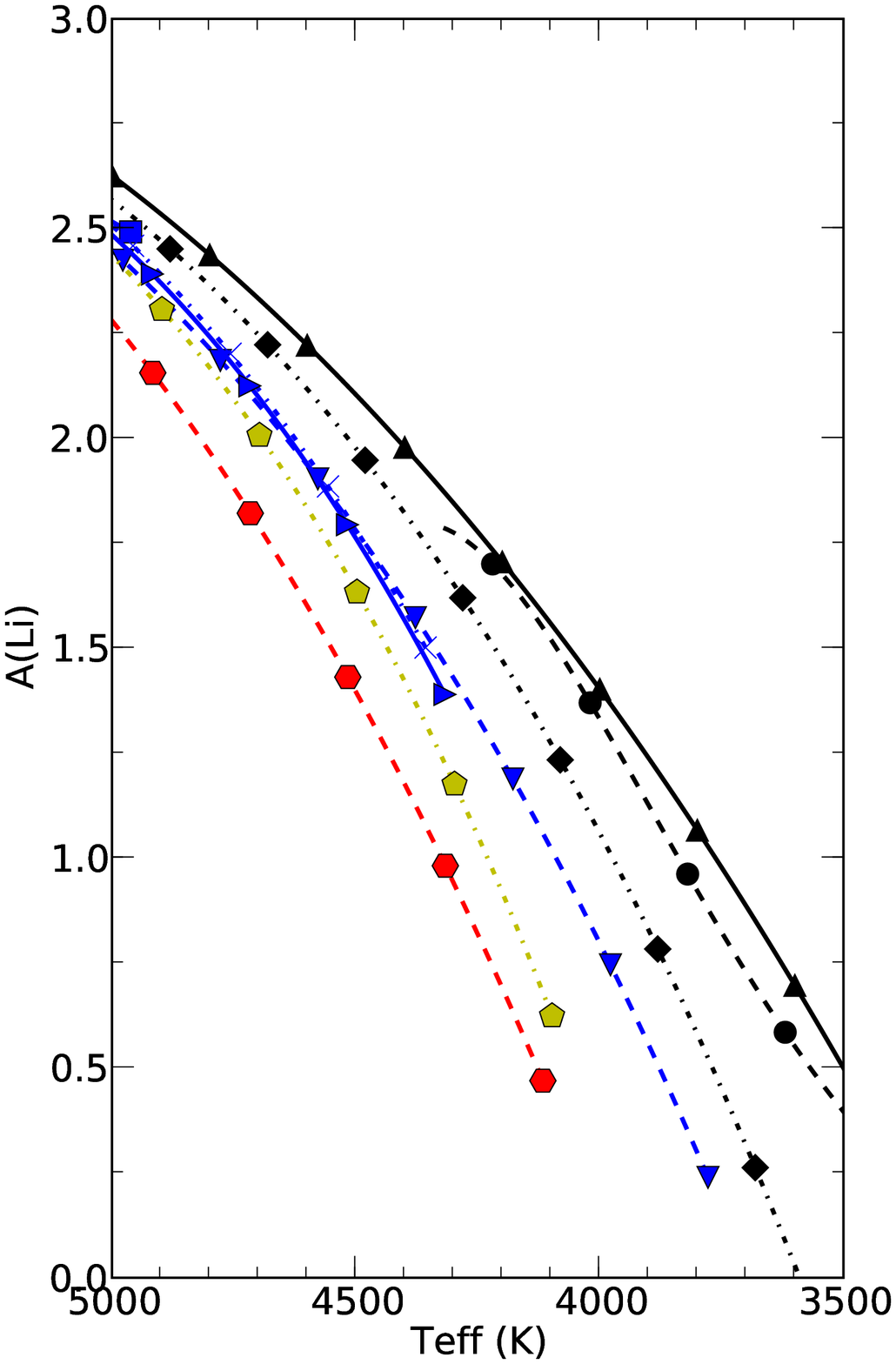}}
\end{center}
\caption{Comparison of the polynomial fit of the observed LDPs. {\it Left.} All polynomial fits for the associations shown in Figure~\ref{fig:ldp}.  
LDPs for each association are identified by a line type and a symbol as follow:
 $\epsilon$ChaA (solid line, filled triangles),TWA (dashed line, filled circles), 
 $\beta$PicA (dotted line, diamonds), OctA (solid line, square), THA (dashed line, downwards triangle), ColA (dashed-dotted line, crosses), CarA (solid line, rightwards triangle), ArgA (dashed-dotted, pentagon) and, ABDA (dashed line, hexagon). The associations between 20-30 Myr are indistinguishable.
 {\it Right.} Zoom in Teff cooler than 4800\.K. At this region a clear separation between the LDP is seen.}
\label{fig:ldp_fit}
\end{figure*}

\begin{figure}
\begin{center}
\resizebox{\hsize}{!}{\includegraphics{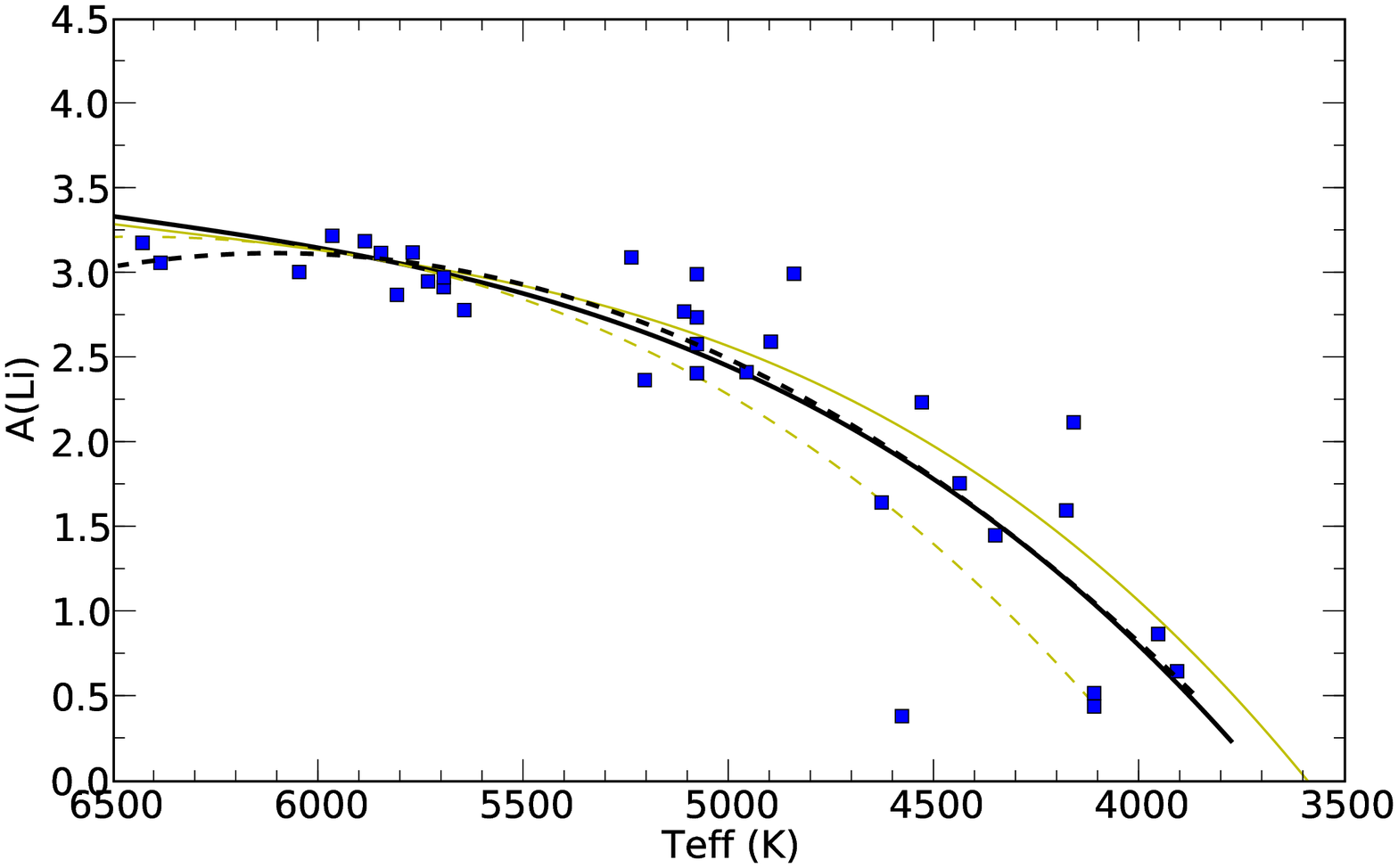}}
\resizebox{\hsize}{!}{\includegraphics{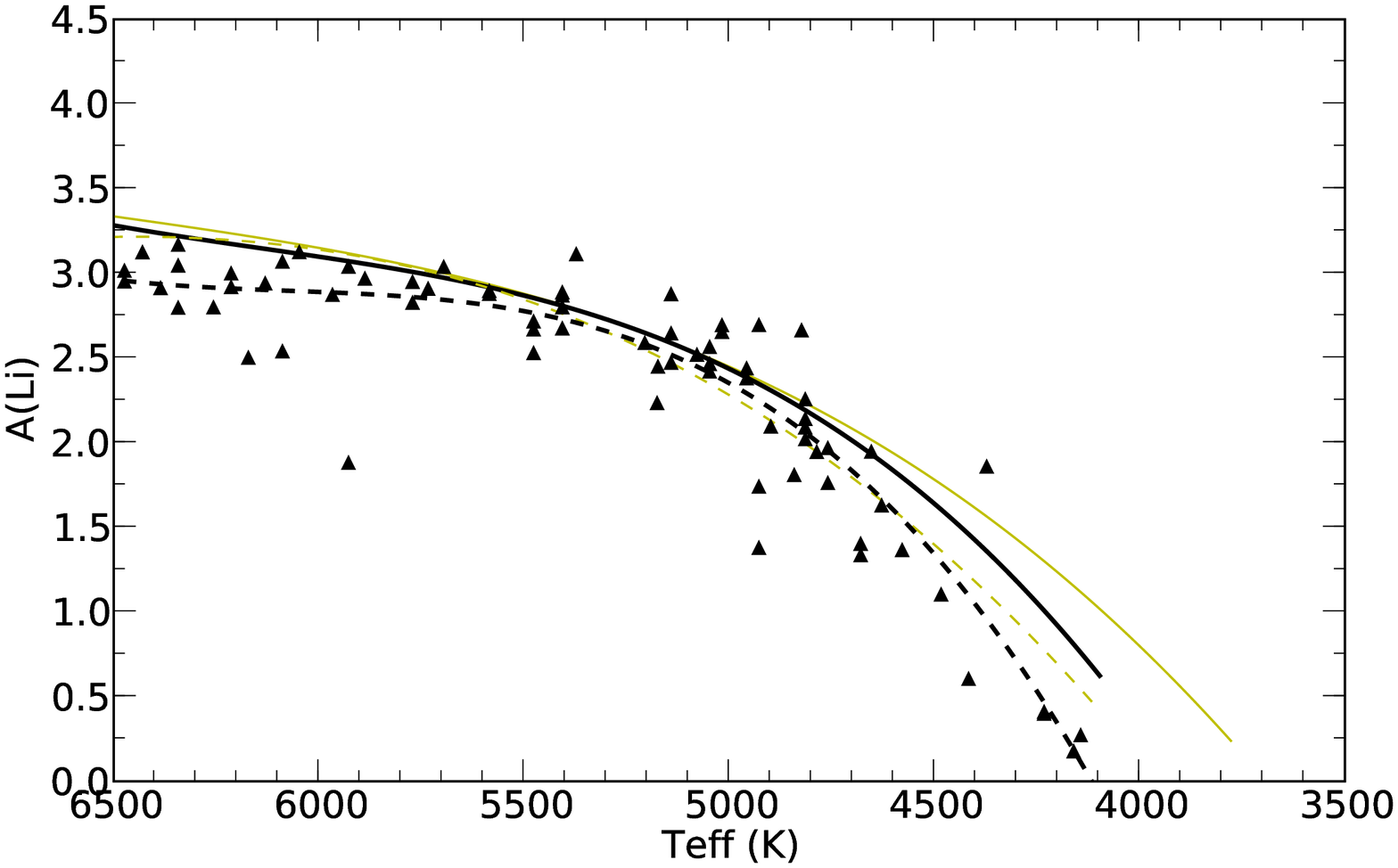}}
\resizebox{\hsize}{!}{\includegraphics{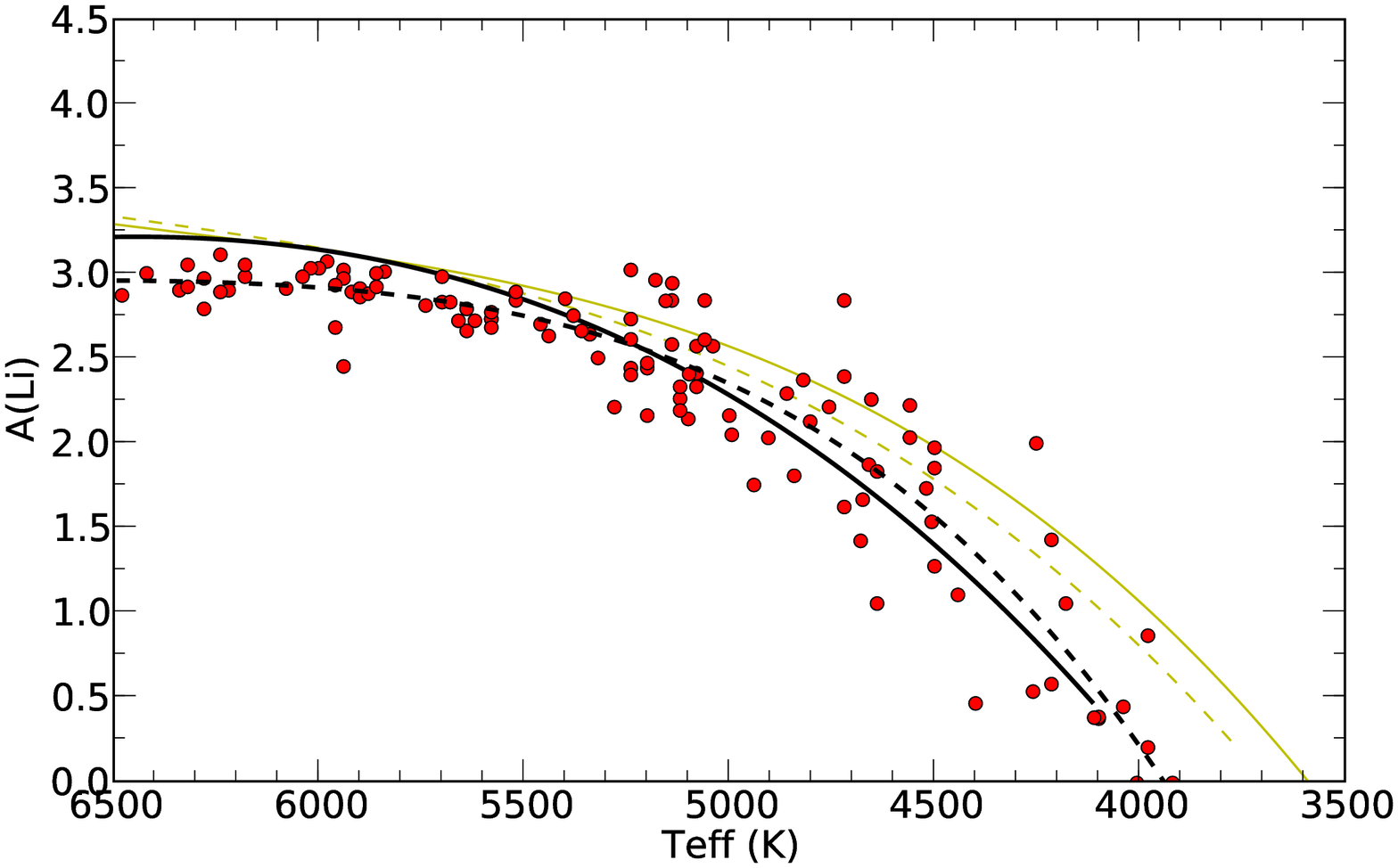}}
\end{center}
\caption{Comparison between the LDPs of the young associations and the Li abundances for the young cluster ($< 100$\,Myr)
from \cite{sestito05}.
{\it Top.} IC~2602 and IC~2391, {\it Middle.} $\alpha$ Per and NGC~2451. {\it Bottom.} The Pleiades.
For each panel our derived LDP having an age closest to that of the cluster is shown as thick black line {\bf (see text)}.
The fitted LDP computed in the same way as for the young associations is shown as thick dashed
black line.}
\label{fig:sestito_us}
\end{figure}

{ 

Using the data from \cite{sestito05}, we computed the LDPs for the young clusters studied by these authors and compared
with those LPDs {\bf of the young associations at similar ages.}

In the top panel of Figure~\ref{fig:sestito_us}, we show the Li abundances for IC~2391 and IC~2602 which have an age
similar to that of THA (30\,Myr). {\bf The LDP of THA is shown as a thick black line} whereas the obtained LDP for these two young clusters are seen as a thick dashed black line. For comparison, the LPD for $\beta$PA (10\,Myr) and ABDA (70\,Myr) 
are shown as light solid and dashed light line, respectively.

In the middle panel of Fig.~\ref{fig:sestito_us}, the Li abundance for $\alpha$ Per and NGC~2451 (50\,Myr) are plotted. In this case, {\bf the LDP shown as thick solid line is that of ArgA.} Solid and dashes light lines are again the LDPs for $\beta$PA and ABDA. Finally, the Li abundances for the Pleiades members are shown in the bottom panel of Fig.~\ref{fig:sestito_us}. The thick black line 
is the LDP of ABDA. The $\beta$PA LDP is shown again as solid light line along with the LDP of THA shown as light dashed line.

{\bf Despite of} the dispersion in the observational data, the agreement between the LDPs of the young clusters
and those of the young association is reasonably good. The exception is the data for $\alpha$ Per and NGC~2451 which show
a level of depletion {\bf close to that of the Pleiades.} According to the LDP of the ArgA, abundances
 $\sim$0.5 dex higher were expected. 

Although the comparison with the young clusters remains marginal (low number of clusters and high dispersion),  {\bf the good agreement} found is already an important result.  
First, it brings confidence in our derivation of the Li abundances described in Sec. 3. 
Secondly, the fact that the LDP of the nine associations are similar to the LDP of open clusters of similar ages
{  strengthens} the notion that the associations presented in Paper\,II are indeed physical groups of stars sharing a common formation history.

}

\subsection{Lithium Depletion Pattern and the relative ages}

All observational LDPs (i.e, the polynomial fits to the observed Li abundances as a function of Teff shown in Figure~\ref{fig:ldp}) 
have been plotted together in the left panel of Figure~\ref{fig:ldp_fit}.  LDPs for each association are 
identified by its line style and marker type (see caption of Figure~\ref{fig:ldp_fit}).  

{  
As expected, Li abundances for stars with temperatures hotter than about 5000\,K are almost constant over the time span ($\sim65$\,Myr) 
covered by our sample of associations \citep{king98,soderblom99,randich97,randich01,stauffer89,martinmontes97,jeffries03,balachandran88,balachandran96,randich98,martin97}. 
}
On the other hand, for those associations possessing members with effective temperatures down to 3600\,K, the observed LDPs 
are clearly distinguishable (right panel of Figure~\ref{fig:ldp_fit}).

A closer look into these different lines in the left panel 
and in the ages quoted in Table~\ref{table:assoc}
shows that indeed there is a gradual shift of the
cool end of the observed LDP as a function of age. 
The age sequence seen in Figure~\ref{fig:ldp_fit} matches the one proposed in Paper\,II based in an isochronal fit of the
color-magnitude diagram restricted to the G- and K-type stars (see Paper\,II for details).
The only exception being the $\epsilon$ChA  which seems to be older 
according to its LDP. 
{  Given the isochronal age of 6\,Myr derived in Paper II, one would expect a flat LDP
around the cosmic Li abundance of $A(Li)= 3.1$ as found in the T-Tauri stars \citep[e.g]{martin94}
and young clusters \citep[e.g.][]{palla05,zapatero02}. The fact that a Li depletion is seen might indicate that $\epsilon$ Cha 
could be a bit older. 
This is discussed below in Sec.~\ref{sec:epsCha}.}

The three associations
with 30\,Myrs, namely, THA, CarA and ColA have been suggested in Paper\,II as being structures of the 
{  Great Austral Young Association (GAYA) } complex. 
From the  Li abundance point of view, these three groups are indistinguishable indicating that indeed they have very similar ages in agreement
with the suggestion of Paper\,II.

From Figure~\ref{fig:ldp},  it is clear that there is an important scatter around the mean LDP for any given associations. 
This scatter  is real and not a  consequence of the errors.
For example,  the stars HD\,6569, HIP\,26401B and UY\,Pic, all
members of the ABDA (that is, with the same age and metallicity) 
have all high-quality observations, similar T$_{\rm eff}$ values 
(therefore similar masses) and $v\sin(i)$ (10, 5 and 9\,km~s$^{-1}$).
However, they show very different A$_{\rm Li}$ values, respectively 2.28, 3.32 and 3.66.

The bottom line is that a distinct  Li depletion history causes an important scatter in the observed LDP preventing the use
of Li as a clock to date individual stars. 
However, statistically speaking, Li abundances derived in a homogeneous way as done in this paper can be used to determine 
relative ages of the young associations 
provided that the associations possess enough members cooler than 5000-4500\,K.
Our conclusions are similar to those of \cite{mentuch08} who also found a qualitatively good agreement between the Li abundances and 
the isochronal ages of a small number of
stars belonging to five associations studied here.

As for the Li "intruders", only two stars (HD\,190102 in the $\beta$PA  and CD-41\,2076 in the ABDA)  
out of the nine rejected
as members of the associations proposed in Paper II based on their low Li abundance
have A(Li) values  relatively close to the LDP of their associations. They
are shown as filled hexagons in Figure~\ref{fig:ldp}. The other are shown only in Tables 4-12.
HD\,190102 was rejected due to the fact that its Li is too low even if the the typical scatter in the Li abundance of the $\beta$PA members is considered.
CD-41\,2076 which has a Li still acceptable for the ABDA lies at 1.1 mag above its isochrone.
{  
In order to either of these two objects to be reconciled as a bonafide member,  their photometric magnitudes (from TYCHO-2) must have a large error and/or they must
be an unresolved binary. In this last case probably the Li abundance could have been underestimated. 
We found no indication suggesting the presence of a companion around these two objects.
In any case, 
these} two examples show that  we must act with caution when eliminating stars based only on Li abundances. 

\subsection{The age of the $\epsilon$Cha Association}
\label{sec:epsCha}
The age estimated in Paper\,II for the $\epsilon$ChA is  6\,Myr which is within the range 
of 3-15\,Myr found in the literature \citep{fernandez08,terranegra99,jilinski05,feigelson03}. 
We should bear in mind  that a given association
might have a different member list according to the method and criteria used to define it. 
Therefore, ages determined  by different methods are not always trivial to be compared. 

The Li abundances for NGC\,2264 (5\,Myr) 
show a flat distribution  around A(Li)$\sim3.2$ for stars with $6500\,K\,>\,T_{eff}\,>\,4000\,K$ 
suggesting that no Li depletion has taken place  \citep{sestito05,king98}. 
\cite{palla05} see no depletion either for the bulk of Orion Nebular Cluster (ONC) (3\,Myr) stars. 
The mean abundance is again 3.1-3.3. 
{ 
Undepleted Lithium abundances were also reported by \cite{zapatero02} for the 
young $\sigma$ Ori cluster. Based on theoretical predictions for the Li depletion, Zapatero Osorio et al. estimated
the age of $\sigma$ Ori to be around 2-4\,Myr.
It is worth noticing that for the ONC and the  $\sigma$ Ori cluster a small group of stars was found to show a considerable Li depletion with respect to the interstellar abundance.  The observed depletion in the Li content was explained 
by \cite{palla05} for the ONC and by \cite{sacco07} for $\sigma$ Ori as a result of an age spread within those two clusters.


Our Figure~\ref{fig:ldp_fit} indeed supports the idea that the age of the $\epsilon$ChaA is at least as young as the TWA but older
 than that of NGC\,2264, ONC and, the $\sigma$ Ori cluster and certainly younger than the $\beta$PA.


}

\subsection{Li and rotation}

\begin{figure}
\begin{center}
\resizebox{\hsize}{!}{\includegraphics[angle=-90]{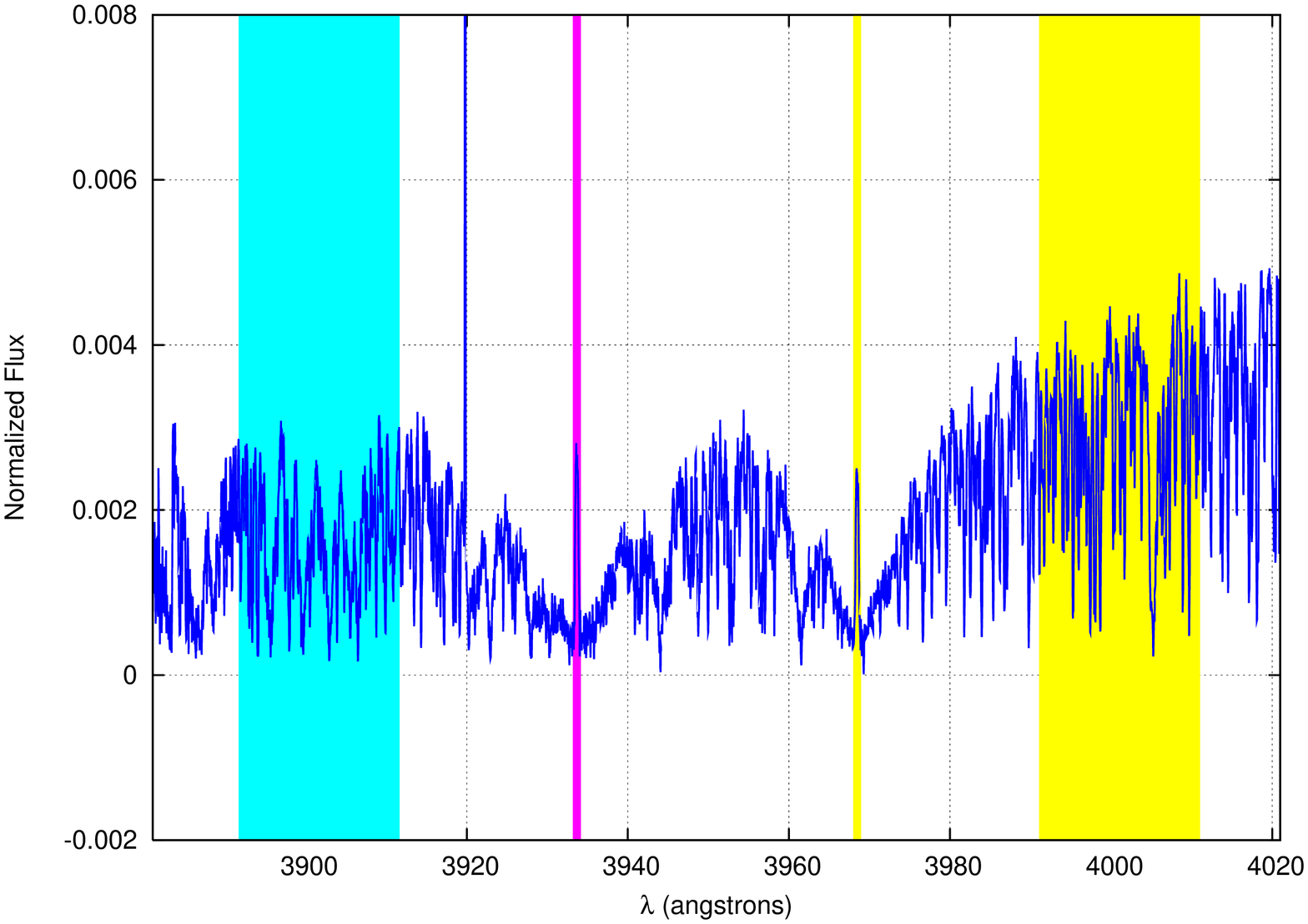}}
\resizebox{\hsize}{!}{\includegraphics[angle=-90]{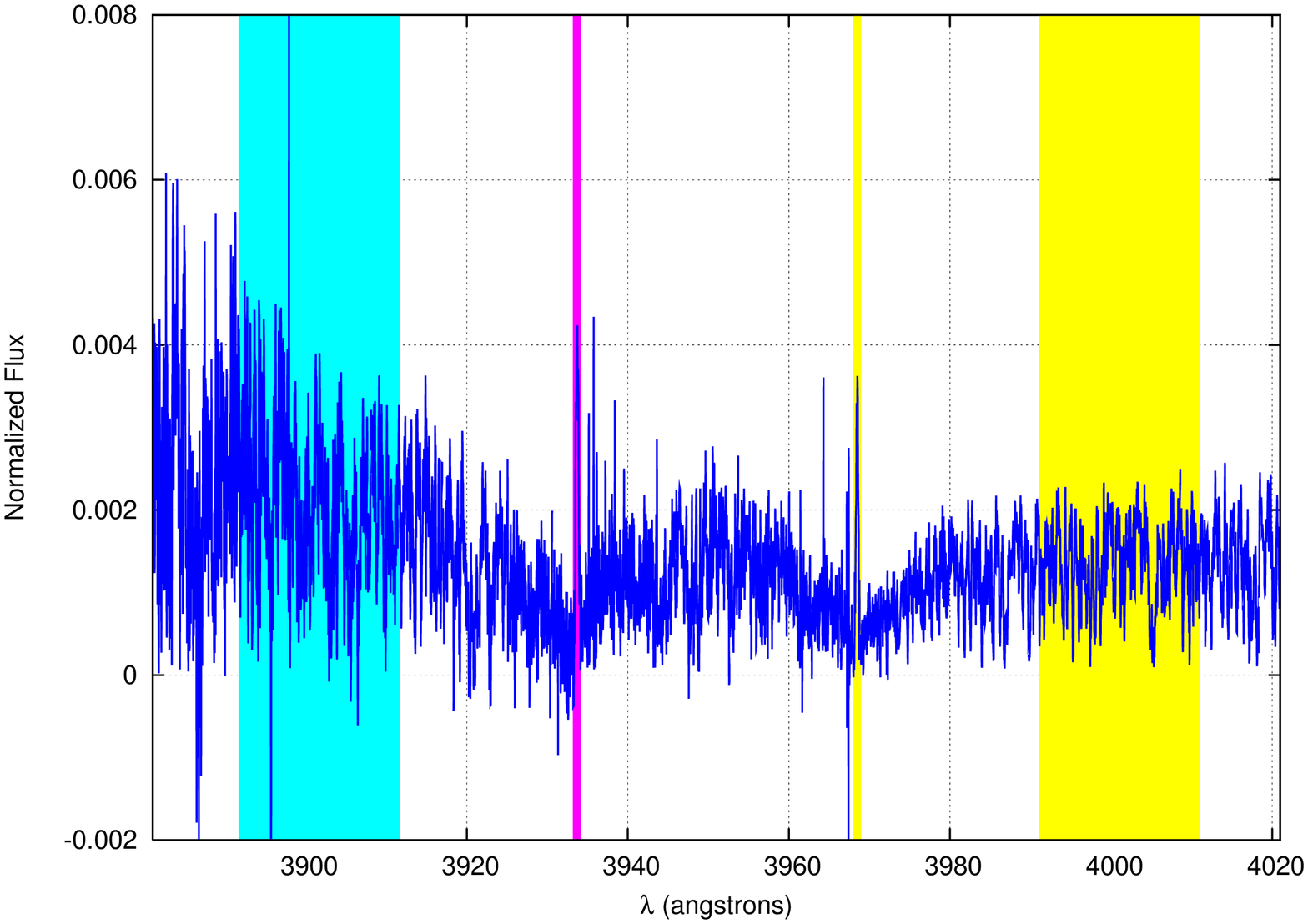}}
\end{center}
\caption{Regions used to compute the CaII H \& K flux. HD~6569 and HIP~26401B are shown in the top and bottom panels,
respectively. }
\label{fig:caii}
\end{figure}

\begin{figure}
\resizebox{\hsize}{!}{\includegraphics{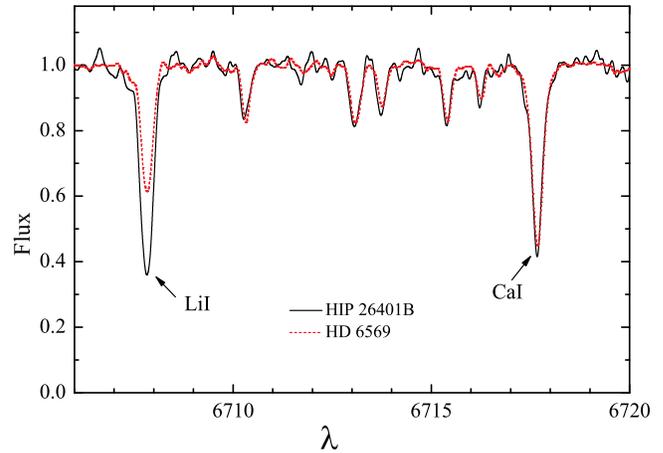}}
\caption{Superposition of the spectra of the stars HD\,6569 and HIP\,26401B in the Li
region. Both stars belong to the AB\,Doradus association and have the same T$_{\rm eff}$ -
note the similarity between the CaI lines, a good temperature indicator - but
have distinct Li line intensities. }
\label{fig:spec_comp}
\end{figure}

\begin{figure}
\resizebox{\hsize}{!}{\includegraphics{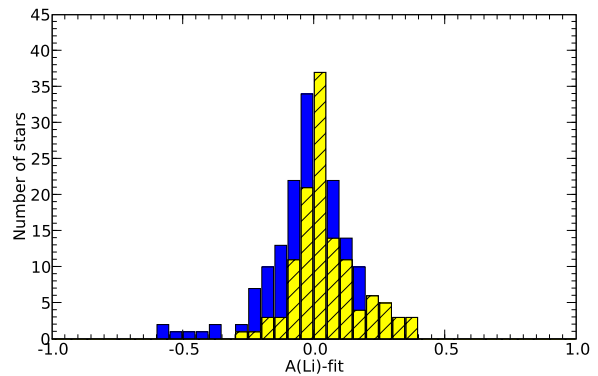}}

\caption{Histogram of the differences between the derived abundances and the polynomial fit of the observed LDP. Dark gray and hatched light gray bins represent
stars rotating slower and faster  than 20\,km\,s$^{-1}$, respectively. }
\label{fig:rot_diff}
\end{figure}

{ 
A careful inspection of Figure~\ref{fig:ldp} shows that stars rotating faster than 20\,km\,s$^{-1}$ 
(filled circles) are often above the polynomial fit of the LDP of the associations suggesting that already at this level, rotation
might play a role in the Li depletion.

This is better seen in Figure~\ref{fig:rot_diff} where
the histograms of the differences between the derived abundances and the polynomial fit of the observed LDP is shown for starts rotating slower and faster than 20 km s$^{-1}$.

Addressing the role of rotation using $V\sin i$ might lead to erroneous conclusions
since the actual rotation of the star is not known due to the $\sin i$ factor.
As an example, we compare in Figure\,\ref{fig:spec_comp}  the spectral region around the Li 6708 line for HD\,6569 ($v\sin i$=10 km s$^{-1}$)and HIP\,26401B ($v\sin i$=5 km s$^{-1}$). {\bf The CaI line at $\lambda\,6718$, a good indicator of temperature  \citep[see][]{cutispoto99}, is also shown in the figure.}
The similitude of their Ca I lines confirms that both stars have very similar T$_{\rm eff}$,
despite the obvious distinct Li line intensities. 

We have used the
rotation-chromospheric flux relation derived by \cite{noyes84}  using the Ca H \& K lines ($R^\prime_{HK}$) to
estimate the rotation period for both stars. The $R^\prime_{HK}$ were derived  as described in \cite{melo06}.
The spectral region around the Ca H \& K lines for both stars is shown in Figure~\ref{fig:caii}.
The calibration of \cite{melo06} yields a $R^\prime_{HK}$ of -4.336 and -4.190 
which translates into a rotation period of 7.2\,days and 2.7\,days for HD~6569 and HIP~26401B, respectively.
According to the  $R^\prime_{HK}$-$P_{rot}$ calibration, HD~6569 is actually rotating almost 3 times slower than HIP~26401B.

Statistically speaking however, Figure~\ref{fig:rot_diff} is worth of mentioning since
the true distribution of equatorial velocities computed from a deconvolution process does not differ considerably
from the projected one \citep[e.g.][]{royer07}.  

There is a vast literature showing that Li depletion is not only driven by convection, 
and that extra-mixing processes (or processes) able to inhibit Li depletion during the PMS) are at work 
\citep[e.g.][]{bouvier08,deliyannis00}. 
The discussion of this complex issue is beyond the scope of this paper. 
{\bf Nevertheless, we point out that Figure~\ref{fig:rot_diff} indicates that a deeper look into the
Li-rotation connection in this sample could be worthwhile.}}



\section{Conclusions}

This is a systematic study of the evolution of the
Li abundances for the most extended sample of pre-main sequence stars belonging to  young, loose,  
nearby associations.
Nine associations with a total of 376 stars have been considered covering ages from $\sim$5\,Myr to
almost that of the age of the Pleiades. 
Our results were compared to  Li studies in young open clusters. 

Our main conclusions are the following:

\begin{itemize}



\item A clear  Li depletion, considered as a measure of a systematic decrease of the
   Li abundance with age, can be measured in the temperature range from
   5000\,K to 3500\,K for the age span covered by the nine associations studied in this paper. 
{ 
\item  The age sequence based on the
Li-clock agrees well with the isochronal ages of Paper\,II.

\item The $\epsilon$ChA being the only possible exception with a LDP showing a considerable Li depletion
for the late type stars in comparison
to young cluster of similar age.
}
\item A real scatter of the Li abundance values, with variations larger than
   those  originating by internal or systematic errors, is present. This scatter hampers the
   use of Li to determine reliable ages for individual stars.

\item The Li depletion patterns for the associations presented here resemble those
of young open clusters with similar ages, strengthening  the notion that the stars of these loose associations have
indeed a common physical origin.


\item For velocities above 20\,km\,s$^{-1}$ rotation seems to play an important role inhibiting the Li depletion. 

\end{itemize}

\begin{acknowledgements}                                             
The authors wish to thank the staff of the Observat\'orio do Pico dos Dias,
LNA/MCT, Brazil and  of the European Southern Observatory, La Silla, Chile.
LS thanks the CNPq, Brazilian Agency, for the grant 301376/86-7.
Sofia Randich is warmly thanked for sharing her Li data with us.
We are grateful to the anonymous referee whose comments helped to improve the quality of the paper.
\end{acknowledgements}

\begin{table*}
\caption{The $\beta$ Pictoris Association}
\begin{tabular}{||lrrrrrr||}
\hline\hline
Ident & $\alpha$(2000) & $\delta$(2000) & EW$_{\rm Li}$ &  T$_{\rm eff}$ & A$_{\rm Li}$&
$v\sin(i)$ \\
~ & ~& ~ & m{\AA} &  K  & ~& km~s$^{-1}$ \\
 \hline
HIP 10679    & 02 17 24.7 & +28 44 30 & 160 &  5988 &  3.08 &   8     \\
HD 14062     & 02 17 25.3 & +28 44 42 & 140 &  6368 &  3.29 &  45     \\
BD+30 397B   & 02 27 28.1 & +30 58 41 & 110 &  3544 & -0.19 &         \\
AG Tri       & 02 27 29.3 & +30 58 25 & 220 &  4351 &  1.55 &   5     \\
BD+05 378    & 02 41 25.9 & +05 59 18 & 450 &  4100 &  1.51 &   9   \\
HD 29391     & 04 37 36.1 & -02 28 25 &     &  7555 &       &  95     \\
GJ 3305      & 04 37 37.5 & -02 29 28 & 120 &  3715 &  0.15 &   5   \\
V1005 Ori    & 04 59 34.8 & +01 47 01 & 270 &  3816 &  0.72 &  14     \\
CD-57 1054   & 05 00 47.1 & -57 15 25 & 360 &  3860 &  0.94 &   6   \\
HIP 23418    & 05 01 58.8 & +09 55 59 &   0 &  3496 &       &   8   \\
BD-21 1074BC & 05 06 49.5 & -21 35 04 &  20 &  3496 & -1.07 &         \\
BD-21 1074A  & 05 06 49.9 & -21 35 09 &  20 &  3613 & -0.88 &         \\
AF Lep       & 05 27 04.8 & -11 54 03 & 191 &  6406 &  3.47 &  52   \\
V1311 Ori    & 05 32 04.5 & -03 05 29 & 100 &  3566 & -0.20 &  12     \\
$\beta$~Pic  & 05 47 17.1 & -51 03 59 &   0 &  8543 &       & 139     \\
AO Men       & 06 18 28.2 & -72 02 41 & 420 &  4377 &  1.91 &  16   \\
HD 139084B   & 15 38 56.8 & -57 42 19 & 460 &  3314 &  0.18 &         \\
V343 Nor     & 15 38 57.5 & -57 42 27 & 292 &  5168 &  2.65 &  17   \\
V824 Ara     & 17 17 25.5 & -66 57 04 & 250 &  5383 &  2.73 &  31     \\
HD 155555C   & 17 17 31.3 & -66 57 06 &  20 &  3413 & -1.17 &   6   \\
GSC 8350-1924 & 17 29 20.7 & -50 14 53&  50 &  3480 & -0.66 &         \\
CD-54 7336   & 17 29 55.1 & -54 15 49 & 360 &  5036 &  2.63 &  35   \\
HD 161460    & 17 48 33.7 & -53 06 43 & 320 &  5140 &  2.62 &  10   \\
HD 164249    & 18 03 03.4 & -51 38 56 & 107 &  6597 &  3.31 &  22   \\
HD 164249B   & 18 03 04.1 & -51 38 56 &  70 &  3607 & -0.30 &         \\
HD 165189    & 18 06 49.9 & -43 25 31 &   0 &  7829 &       & 104     \\
V4046 Sgr    & 18 14 10.5 & -32 47 33 & 440 &  4361 &  1.92 &  14   \\
GSC 7396-0759 & 18 14 22.1 & -32 46 10& 190 &  3619 &  0.18 &   3   \\
HD 168210    & 18 19 52.2 & -29 16 33 & 290 &  5645 &  3.10 & 115   \\
HD 172555    & 18 45 26.9 & -64 52 17 &   0 &  8323 &       & 116     \\
CD-64 1208   & 18 45 36.9 & -64 51 48 & 490 &  4148 &  1.65 & 110   \\
TYC 9073-0762-1 & 18 46 52.6&-62 10 36& 332 &  3649 &  0.49 &   10   \\
CD-31 16041  & 18 50 44.5 & -31 47 47 & 492 &  3889 &  1.17 &  50   \\
PZ Tel       & 18 53 05.9 & -50 10 50 & 287 &  5344 &  2.77 &  69  \\
TYC 6872-1011-1 & 18 58 04.2&-29 53 05& 483 &  3850 &  1.08 &  34   \\
CD-26 13904  & 19 11 44.7 & -26 04 09 & 320 &  4655 &  2.13 &   10   \\
$\eta$ Tel   & 19 22 51.2 & -54 25 26 &   0 &  9670 &       & 330     \\
HD 181327    & 19 22 58.9 & -54 32 17 & 120 &  6597 &  3.36 &  21   \\
HD 191089    & 20 09 05.2 & -26 13 26 &  95 &  6521 &  3.21 &  45     \\
AT MicB      & 20 41 51.1 & -32 26 10 &   0 &  3331 &       &  16   \\
AT MicA      & 20 41 51.2 & -32 26 07 &   0 &  3341 &       &  10   \\
AU Mic       & 20 45 09.5 & -31 20 27 &  80 &  3649 & -0.16 &   9   \\
HD 199143    & 20 55 47.7 & -17 06 51 & 147 &  6330 &  3.29 & 129   \\
AZ Cap       & 20 56 02.7 & -17 10 54 & 420 &  4197 &  1.64 &  16   \\
CP-72 2713   & 22 42 48.9 & -71 42 21 & 440 &  3921 &  1.17 &   8   \\
WW PsA       & 22 44 58.0 & -33 15 02 &   0 &  3391 &       &  12   \\
TX PsA       & 22 45 00.0 & -33 15 26 & 450 &  3316 & -0.09 &  17   \\
BD-13 6424   & 23 32 30.9 & -12 15 52 & 185 &  3715 &  0.35 &   9   \\
\hline
\multicolumn{7}{||c||} {Intruders}\\
\hline	
HD 190102   &20 04 18.1 & -26  19 46 & 110 &   5874 &  2.82     &      \\
TYC 9114-1267-1& 21 21 28.7 & -66  55 06 &  15 & 3889 & -0.47      &   4 \\
\hline
\hline                                                    
\end{tabular}                                                       
\end{table*}  
\begin{table*}
\caption{The Tucana-Horologium Association}
\begin{tabular}{||lrrrrrr||}
\hline\hline
Ident & $\alpha$(2000) & $\delta$(2000) & EW$_{\rm Li}$ & T$_{\rm eff}$ & A$_{\rm Li}$& $v\sin(i)$ \\
~ & ~& ~ & m{\AA} &  K  & ~& km~s$^{-1}$ \\
\hline
HD 105       & 00 05 52.5 & -41 45 11 & 153 &  6178 &  3.20 &  14   \\
HD 987       & 00 13 53.0 & -74 41 18 & 202 &  5481 &  2.77 &   7   \\
HD 1466      & 00 18 26.1 & -63 28 39 & 130 &  6368 &  3.26 &  18   \\
HIP 1910     & 00 24 09.0 & -62 11 04 & 194 &  3860 &  0.65 &  21   \\
CT Tuc       & 00 25 14.7 & -61 30 48 &  40 &  3841 & -0.13 &   7   \\
HD 2884      & 00 31 32.7 & -62 57 30 &   0 &  9670 &       & 107     \\
HD 2885      & 00 31 33.5 & -62 57 56 &  18 &  9036 &  3.90 &   6   \\
HD 3003      & 00 32 43.9 & -63 01 53 &   0 &  9146 &       &  78     \\
HD 3221      & 00 34 51.2 & -61 54 58 & 360 &  4309 &  1.73 & 123   \\
CD-78 24     & 00 42 20.3 & -77 47 40 & 291 &  4540 &  1.94 &  19   \\
HD 8558      & 01 23 21.3 & -57 28 51 & 196 &  5683 &  2.94 &  14   \\
CC Phe       & 01 28 08.7 & -52 38 19 & 168 &  4894 &  2.09 &   3   \\
DK Cet       & 01 57 49.0 & -21 54 05 & 190 &  5950 &  3.14 &  15   \\
HD 13183     & 02 07 18.1 & -53 11 56 & 229 &  5721 &  3.05 &  24   \\
HD 13246     & 02 07 26.1 & -59 40 46 & 141 &  6368 &  3.30 &  36   \\
CD-60 416    & 02 07 32.2 & -59 40 21 & 254 &  4268 &  1.50 &  11   \\
$\phi$ Eri   & 02 16 30.6 & -51 30 44 &   0 & 15665 &       & 240     \\
$\epsilon$ Hyi & 02 39 35.4&-68 16 01 &   0 & 10852 &       &  96     \\
CD-53 544    & 02 41 46.8 & -52 59 52 & 298 &  4087 &  1.27 &  80   \\
AF Hor       & 02 41 47.3 & -52 59 31 &  10 &  3511 & -1.35 &  10   \\
CD-58 553    & 02 42 33.0 & -57 39 37 & 120 &  4216 &  1.07 &   6   \\
CD-35 1167   & 03 19 08.7 & -35 07 00 &  65 &  4100 &  0.59 &   6   \\
CD-46 1064   & 03 30 49.1 & -45 55 57 & 229 &  4487 &  1.75 &         \\
CD-44 1173   & 03 31 55.7 & -43 59 14 & 251 &  4110 &  1.23 &  11   \\
HD 22213     & 03 34 16.4 & -12 04 07 & 260 &  5729 &  3.12 &     42    \\
HD 22705     & 03 36 53.4 & -49 57 29 & 154 &  6254 &  3.26 &  18   \\
BD-12 943    & 04 36 47.1 & -12 09 21 & 240 &  5460 &  2.84 &       17  \\
HD 29615     & 04 38 43.9 & -27 02 02 & 200 &  6026 &  3.22 &  18   \\
HD 30051     & 04 43 17.2 & -23 37 42 & 120 &  6798 &  3.44 &         \\
TYC 8083-0455-1 & 04 48 00.7&-50 41 26 &  40 &  3908 &  0.01 &   5   \\
HD 32195     & 04 48 05.2 & -80 46 45 & 130   &  6368 & 3.26  &  41   \\
BD-20 951    & 04 52 49.5 & -19 55 02 & 190 &  5083 &  2.35 &         \\
BD-19 1062   & 04 59 32.0 & -19 17 42 & 230 &  4735 &  2.06 &   12      \\
BD-09 1108   & 05 15 36.5 & -09 30 51 & 245 &  5683 &  3.05 &  18     \\
CD-30 2310   & 05 18 29.0 & -30 01 32 & 310 &  4158 &  1.42 &   7   \\
HD 53842     & 06 46 13.5 & -83 59 30 &    &  6597 &       &        \\
$\alpha$ Pav & 20 25 38.9 & -56 44 06 &   0 & 17726 &       &  35     \\
HD 202917    & 21 20 50.0 & -53 02 03 & 227 &  5567 &  2.91 &  15   \\
HIP 107345   & 21 44 30.1 & -60 58 39 &  55 &  3824 & -0.01 &   8   \\
HD 207575    & 21 52 09.7 & -62 03 09 & 110 &  6483 &  3.25 &  30   \\
HD 207964    & 21 55 11.4 & -61 53 12 & 100 &  7006 &  3.48 & 110     \\
TYC 9344-0293-1 & 23 26 10.7 & -73 23 50 & 123 &  3816 &  0.36 &  61   \\
CD-86 147    & 23 27 49.4 & -86 13 19 & 276 &  5481 &  2.88 &  74   \\
HD 222259B   & 23 39 39.3 & -69 11 40 & 232 &  4629 &  1.94 &  15   \\
DS Tuc       & 23 39 39.5 & -69 11 45 & 216 &  5683 &  2.98 &  18   \\
\hline
\multicolumn{7}{||c||} {Intruders}\\
\hline	
CD-34 521   &01 22 04.4 & -33  37  04&   0 &   4226 &       &   5  \\
\hline\hline	                                                    
\end{tabular}                                                       
\end{table*}                                           
\begin{table*}
\caption{The Columba Association}
\begin{tabular}{||lrrrrrr||}
\hline\hline
Ident & $\alpha$(2000) & $\delta$(2000) & EW$_{\rm Li}$ &  T$_{\rm eff}$ & A$_{\rm Li}$& $v\sin(i)$ \\
~ & ~& ~ & m{\AA} &  K  & ~& km~s$^{-1}$ \\
\hline
BD-16 351    & 02 01 35.6 & -16 10 01 & 190 &  5083 &  2.36 &    9      \\
BD-11 648    & 03 21 49.7 & -10 52 18 & 320 &  5225 &  2.77 &      27    \\
V1221 Tau    & 03 28 15.0 & +04 09 48 & 275 &  5645 &  3.08 &  96      \\
HD 21955     & 03 31 20.8 & -30 30 59 & 230 &  5835 &  3.10 &  22    \\
HD 21997     & 03 31 53.6 & -25 36 51 &   0 &  8707 &       &  70      \\
BD-04 700    & 03 57 37.2 & -04 16 16 & 250 &  5506 &  2.91 &      14    \\
BD-15 705    & 04 02 16.5 & -15 21 30 & 220 &  4735 &  2.04 &   7    \\
HD 26980     & 04 14 22.6 & -38 19 02 & 183 &  6140 &  3.27 &  14    \\
HD 27679     & 04 21 10.3 & -24 32 21 & 180 &  5928 &  3.10 &          \\
CD-43 1395   & 04 21 48.7 & -43 17 33 & 270 &  5683 &  3.10 &  26    \\
CD-36 1785   & 04 34 50.8 & -35 47 21 & 300 &  4941 &  2.43 &   9    \\
BD+08 742    & 04 42 32.1 & +09 06 01 & 240 &  5225 &  2.62 &    14      \\
HD 30447     & 04 46 49.5 & -26 18 09 &    &  6897 &       &  70      \\
GSC 8077-1788 & 04 51 53.0 & -46 47 31 &  30 &  3655 & -0.62 &          \\
HD 31242     & 04 51 53.5 & -46 47 13 & 250 &  5721 &  3.09 &          \\
HD 272836    & 04 53 05.2 & -48 44 39 & 250 &  4917 &  2.31 &          \\
TYC 5900-1180-1 & 04 58 35.8 & -15 37 31 & 290 &  5383 &  2.87 &  40      \\
BD-08 995    & 04 58 48.6 & -08 43 40 & 270 &  5225 &  2.68 &       7   \\
HD 32372     & 05 00 51.9 & -41 01 07 & 210 &  5721 &  3.01 &   8    \\
AS Col       & 05 20 38.0 & -39 45 18 & 140 &  6483 &  3.37 &  72    \\
BD-08 1115   & 05 24 37.3 & -08 42 02 & 280 &  5432 &  2.90 &  33      \\
HD 35841     & 05 26 36.6 & -22 29 24 &    &  6556 &       &          \\
HD 274561    & 05 28 55.1 & -45 34 58 & 270 &  4781 &  2.20 &   7    \\
HD 36329     & 05 29 24.1 & -34 30 56 & 220 &  5912 &  3.19 &   8    \\
AG Lep       & 05 30 19.1 & -19 16 32 & 230 &  5683 &  3.02 &  21      \\
HD 37484     & 05 37 39.6 & -28 37 35 & 125 &  6848 &  3.53 &  43      \\
TYC 0119-1242-1 & 05 37 45.3 & +02 30 57 & 330 &  4361 &  1.76 &    10      \\
TYC 0119-0497-1 & 05 37 46.5 & +02 31 26 & 300 &  4361 &  1.72 &     13     \\
BD-08 1195   & 05 38 35.0 & -08 56 40 & 290 &  5352 &  2.84 &  29      \\
HD 38207     & 05 43 21.0 & -20 11 21 &    &  6656 &       &          \\
HD 38206     & 05 43 21.7 & -18 33 27 &   0 &  9366 &       &  41      \\
CD-38 2198   & 05 45 16.3 & -38 36 49 & 250 &  5481 &  2.88 &  22    \\
CD-29 2531   & 05 50 21.4 & -29 15 21 & 270 &  5225 &  2.68 &          \\
CD-52 1363   & 05 51 01.2 & -52 38 13 & 290 &  5225 &  2.66 &  53    \\
HD 40216     & 05 55 43.2 & -38 06 16 & 130 &  6483 &  3.34 &  33    \\
V1358 Ori    & 06 19 08.1 & -03 26 20 & 170 &  6178 &  3.26 &  36    \\
CD-40 2458   & 06 26 06.9 & -41 02 54 & 310 &  5314 &  2.84 &  12    \\
CD-48 2324   & 06 28 06.1 & -48 26 53 & 280 &  5383 &  2.85 &  41    \\
TYC 4810-0181-1 & 06 31 55.2 & -07 04 59 & 250 &  4735 &  2.11 & 125      \\
HD 295290    & 06 40 22.3 & -03 31 59 & 330 &  5314 &  2.87 &  39      \\
HD 48370     & 06 43 01.0 & -02 53 19 & 170 &  5663 &  2.85 &   9      \\
CD-36 3202   & 06 52 46.7 & -36 36 17 & 300 &  4917 &  2.41 & 170    \\
HD 51797     & 06 56 23.5 & -46 46 55 & 335 &  5390 &  2.95 &  16    \\
HD 62237     & 07 42 26.6 & -16 17 00 & 230 &  6127 &  3.37 &          \\
\hline\hline	
\end{tabular}  
\end{table*}  
\begin{table*}
\caption{The Carina  Association}
\begin{tabular}{||lrrrrrr||}
\hline\hline
Ident & $\alpha$(2000) & $\delta$(2000) & EW$_{\rm Li}$ & T$_{\rm eff}$ & A$_{\rm Li}$& $v\sin(i)$ \\
~ & ~& ~ & m{\AA} &  K  & ~& km~s$^{-1}$ \\
\hline
HD 42270     & 05 53 29.3 & -81 56 53 & 305 &  4988 &  2.49 &  30   \\
AB Pic       & 06 19 12.9 & -58 03 16 & 320 &  5168 &  2.70 &  12   \\
HD 49855     & 06 43 46.2 & -71 58 35 & 233 &  5721 &  3.05 &  12   \\
HD 55279     & 07 00 30.5 & -79 41 46 & 278 &  5036 &  2.49 &   9   \\
CD-57 1709   & 07 21 23.7 & -57 20 37 & 247 &  5225 &  2.63 &  12   \\
CD-63 408    & 08 24 06.0 & -63 34 03 & 205 &  5759 &  3.02 &  73   \\
CD-61 2010   & 08 42 00.5 & -62 18 26 & 275 &  5140 &  2.59 &  38   \\
CD-53 1875   & 08 45 52.7 & -53 27 28 & 170 &  5931 &  3.07 &  45   \\
CD-75 392    & 08 50 05.4 & -75 54 38 & 261 &  5481 &  2.90 &  44   \\
CD-53 2515   & 08 51 56.4 & -53 55 57 & 240 &  5607 &  2.97 &  29   \\
TYC 8582-3040-1 & 08 57 45.6&-54 08 37& 320 &  4735 &  2.16 &  24   \\
CD-49 4008   & 08 57 52.2 & -49 41 51 & 260 &  5567 &  2.98 &  33   \\
CD-54 2499   & 08 59 28.7 & -54 46 49 & 240 &  5759 &  3.05 & 128   \\
CP-55 1885   & 09 00 03.4 & -55 38 24 & 250 &  5645 &  3.03 &  44   \\
CD-55 2543   & 09 09 29.4 & -55 38 27 & 200 &  5506 &  2.79 &  15   \\
CD-54 2644   & 09 13 16.9 & -55 29 03 & 240 &  5607 &  2.97 &  40   \\
V479 Car     & 09 23 35.0 & -61 11 36 & 345 &  5012 &  2.58 &  15   \\
HD 83096     & 09 31 24.9 & -73 44 49 & 100 &  7116 &  3.57 &  53   \\
HIP 46720B   & 09 31 25.2 & -73 44 51 & 240 &  5383 &  2.77 &         \\
CP-52 2481   & 09 32 26.1 & -52 37 40 & 245 &  5197 &  2.59 &  17   \\
CP-62 1293   & 09 43 08.8 & -63 13 04 & 215 &  5481 &  2.80 &  35   \\
CD-54 4320   & 11 45 51.8 & -55 20 46 & 190 &  4361 &  1.50 &   5   \\
HD 107722    & 12 23 29.0 & -77 40 51 &  80 &  6556 &  3.15 &  30     \\
\hline
\multicolumn{7}{||c||} {Intruders}\\
\hline
CD-48 4797  &09 33 14.3 & -48  48 33 &   0 &   4226 &       &  80  \\
\hline\hline	                                                    
\end{tabular}   
\end{table*}

\begin{table*}
\caption{The TW~Hydrae Association}
\begin{tabular}{||lrrrrrr||}
\hline\hline
Ident & $\alpha$(2000) & $\delta$(2000) & EW$_{\rm Li}$ &  T$_{\rm eff}$ & A$_{\rm Li}$& $v\sin(i)$ \\
~ & ~& ~ & m{\AA} &  K  & ~& km~s$^{-1}$ \\
\hline
TWA 7        & 10 42 30.1 & -33 40 17 & 530 &  3503 &  0.49 &   4   \\
TWA 1        & 11 01 51.9 & -34 42 17 & 435 &  3973 &  1.26 &   6   \\
TWA 2        & 11 09 13.8 & -30 01 40 & 535 &  3649 &  0.74 &  13   \\
TWA 3B       & 11 10 27.8 & -37 31 53 & 580 &  3321 &  0.31 &  12   \\
TWA 3A       & 11 10 27.9 & -37 31 52 & 710 &  3376 &  0.49 &  12   \\
TWA 13A      & 11 21 17.2 & -34 46 46 & 580 &  3779 &  1.04 &  12   \\
TWA 13B      & 11 21 17.4 & -34 46 50 & 550 &  3655 &  0.77 &  12     \\
TWA 4        & 11 22 05.3 & -24 46 40 & 380 &  4299 &  1.74 &   9   \\
TWA 5A       & 11 31 55.3 & -34 36 27 & 629 &  3533 &  0.63 &  54   \\
TWA 5B       & 11 31 55.4 & -34 36 29 & 300 &  3050 & -0.23 &  16     \\
TWA 8A       & 11 32 41.2 & -26 51 56 & 600 &  3514 &  0.58 &   7   \\
TWA 8B       & 11 32 41.2 & -26 52 09 & 560 &  3240 &  0.21 &  11   \\
TWA 26       & 11 39 51.1 & -31 59 21 & 500 &  3050 &  0.03 &  25     \\
TWA 9B       & 11 48 23.7 & -37 28 48 & 480 &  3458 &  0.37 &   8   \\
TWA 9A       & 11 48 24.2 & -37 28 49 & 470 &  4062 &  1.47 &  11   \\
TWA 27       & 12 07 33.4 & -39 32 54 & 500 &  3107 &  0.06 &  13     \\
TWA 25       & 12 15 30.7 & -39 48 43 & 555 &  3742 &  0.94 &  13   \\
TWA 20       & 12 31 38.1 & -45 58 59 & 160 &  3492 & -0.10 &  30     \\
TWA 16       & 12 34 56.4 & -45 38 07 & 360 &  3649 &  0.53 &  11     \\
TWA 10       & 12 35 04.2 & -41 36 39 & 500 &  3492 &  0.44 &   6   \\
TWA 11B      & 12 36 00.6 & -39 52 16 & 550 &  3592 &  0.65 &  12   \\
TWA 11A      & 12 36 01.0 & -39 52 10 &   0 &  9311 &       & 152     \\
\hline\hline	                                                     
\end{tabular}                                                        
\end{table*}
\begin{table*}
\caption{The $\epsilon$ Chamaleontis Association}
\begin{tabular}{||lrrrrrr||}
\hline\hline
Ident & $\alpha$(2000) & $\delta$(2000) & EW$_{\rm Li}$ &  T$_{\rm eff}$ & A$_{\rm Li}$& $v\sin(i)$ \\
~ & ~& ~ & m{\AA} &  K  & ~& km~s$^{-1}$ \\
\hline
EG Cha       & 08 36 56.2 & -78 56 46 & 510 &  4268 &  1.86 &  22   \\
$\eta$~Cha   & 08 41 19.5 & -78 57 48 &   0 & 11384 &       & 390     \\
RS Cha       & 08 43 12.2 & -79 04 12 &   0 &  8049 &       &  64     \\
EQ Cha       & 08 47 56.8 & -78 54 53 & 570 &  3529 &  0.57 &  15   \\
CP-68 1388   & 10 57 49.4 & -69 14 00 & 420 &  4695 &  2.33 &  26   \\
DZ Cha       & 11 49 31.9 & -78 51 01 & 560 &  3603 &  0.68 &  18     \\
T Cha        & 11 57 13.5 & -79 21 32 & 360 &  5111 &  2.71 &  39     \\
GSC 9415-2676 & 11 58 26.9 & -77 54 45 & 600 &  3486 &  0.53 &   5     \\
EE Cha       & 11 58 35.2 & -77 49 31 &   0 &  8104 &       &  93     \\
$\epsilon$ Cha & 11 59 37.6& -78 13 19&   0 & 11384 &       & 265     \\
HIP 58490    & 11 59 42.3 & -76 01 26 & 445 &  4351 &  1.91 &  10   \\
DX Cha       & 12 00 05.1 & -78 11 35 &   0 &  7994 &       &  12     \\
HD 104237D   & 12 00 08.3 & -78 11 40 & 580 &  3434 &  0.44 &  6      \\
HD 104237E   & 12 00 09.3 & -78 11 42 & 480 &  3850 &  1.08 &  30       \\
HD 104467    & 12 01 39.1 & -78 59 17 & 260 &  5759 &  3.14 &  22   \\
GSC 9420-0948 & 12 02 03.8 & -78 53 01 & 540 &  3661 &  0.77 &  12     \\
GSC 9416-1029 & 12 04 36.2 & -77 31 35 & 470 &  3537 &  0.47 &   6     \\
HD 105923    & 12 11 38.1 & -71 10 36 & 280 &  5344 &  2.81 &  13   \\
GSC 9239-1495 & 12 19 43.8 & -74 03 57 & 560 &  3673 &  0.81 &         \\
GSC 9239-1572 & 12 20 21.9 & -74 07 39 & 610 &  3749 &  1.01 &  41     \\
CD-74 712    & 12 39 21.3 & -75 02 39 & 459 &  4722 &  2.41 &  20   \\
CD-69 1055   & 12 58 25.6 & -70 28 49 & 400 &  4917 &  2.56 &  24   \\
MP Mus       & 13 22 07.5 & -69 38 12 & 424 &  4616 &  2.24 &  14   \\
\hline\hline	                                                    
\end{tabular}                                                       
\end{table*}  
\begin{table*}
\caption{The Octans Association}
\begin{tabular}{||lrrrrrr||}
\hline\hline
Ident & $\alpha$(2000) & $\delta$(2000) & EW$_{\rm Li}$ &  T$_{\rm eff}$ & A$_{\rm Li}$& $v\sin(i)$ \\
~ & ~& ~ & m{\AA} &  K  & ~& km~s$^{-1}$ \\
\hline
CD-58 860    & 04 11 55.7 & -58 01 47 & 225 &  5759 &  3.07 &  20   \\
CD-43 1451   & 04 30 27.3 & -42 48 47 & 280 &  5168 &  2.63 &  19   \\
CD-72 248    & 05 06 50.6 & -72 21 12 & 350 &  5059 &  2.58 & 190   \\
HD 274576    & 05 28 51.4 & -46 28 19 & 235 &  5683 &  3.03 &  21   \\
CD-47 1999   & 05 43 32.1 & -47 41 11 & 190 &  6064 &  3.23 &  39   \\
TYC 7066-1037-1 & 05 58 11.8 & -35 00 49 & 250 &  5383 &  2.79 &  20   \\
CD-66 395    & 06 25 12.4 & -66 29 10 & 250 &  5225 &  2.57 & 190   \\
CD-30 3394A  & 06 40 04.9 & -30 33 03 & 120 &  6490 &  3.30 &  36   \\
CD-30 3394B  & 06 40 05.7 & -30 33 09 & 170 &  6140 &  3.23 &  40   \\
HD 155177    & 17 42 09.0 & -86 08 05 &  80 &  6623 &  3.19 &  21   \\
TYC 9300-0529-1 & 18 49 45.1 & -71 56 58 & 300 &  5140 &  2.64 &  26   \\
TYC 9300-0891-1 & 18 49 48.7 & -71 57 10 & 310 &  5168 &  2.69 &   9   \\
CP-79 1037   & 19 47 03.9 & -78 57 43 & 257 &  5304 &  2.73 &  29   \\
CP-82 784    & 19 53 56.8 & -82 40 42 & 275 &  5012 &  2.46 &  39   \\
CD-87 121    & 23 58 17.7 & -86 26 24 & 266 &  5225 &  2.66 &  34   \\
\hline\hline	                                                    
\end{tabular}                                                       
\end{table*} 

\begin{table*}
\caption{The Argus Association}
\begin{tabular}{||lrrrrrr||}
\hline\hline
Ident & $\alpha$(2000) & $\delta$(2000) & EW$_{\rm Li}$ &  T$_{\rm eff}$ & A$_{\rm Li}$& $v\sin(i)$ \\
~ & ~& ~ & m{\AA} &  K  & ~& km~s$^{-1}$ \\
\hline
BW Phe       & 00 56 55.5 & -51 52 32 & 148 &  4826 &  1.96 &  28   \\
CD-49 1902   & 05 49 44.8 & -49 18 26 & 220 &  5645 &  2.96 &  55   \\
CD-56 1438   & 06 11 53.0 & -56 19 05 & 230 &  5225 &  2.59 & 130   \\
CD-28 3434   & 06 49 45.4 & -28 59 17 & 230 &  5607 &  2.95 &         \\
CD-42 2906   & 07 01 53.4 & -42 27 56 & 275 &  5083 &  2.54 &  11   \\
CD-48 2972   & 07 28 22.0 & -49 08 38 & 250 &  5506 &  2.90 &  52   \\
CD-48 3199   & 07 47 26.0 & -49 02 51 & 230 &  5607 &  2.95 &  25   \\
CD-43 3604   & 07 48 49.8 & -43 27 06 & 320 &  4589 &  2.05 &  40   \\
TYC 8561-0970-1 & 07 53 55.5&-57 10 07& 210 &  5225 &  2.55 &   6   \\
HD 67945     & 08 09 38.6 & -20 13 50 &   0 &  7159 &       & 120   \\
CD-58 2194   & 08 39 11.6 & -58 34 28 & 270 &  5759 &  3.16 &  85   \\
PMM 7422     & 08 28 45.6 & -52 05 27 & 233 &  5683 &  3.02 &  33   \\
PMM 7956     & 08 29 51.9 & -51 40 40 & 289 &  4735 &  2.17 &  14   \\
PMM 1560     & 08 29 52.4 & -53 22 00 & 107 &  5874 &  2.80 &   6   \\
PMM 6974     & 08 34 18.1 & -52 15 58 &  74 &  4633 &  1.42 &   5   \\
PMM 4280     & 08 34 20.5 & -52 50 05 & 151 &  5759 &  2.87 &  16   \\
PMM 6978     & 08 35 01.2 & -52 14 01 & 179 &  4667 &  1.86 &   7   \\
PMM 2456     & 08 35 43.7 & -53 21 20 & 301 &  4735 &  2.20 &  48   \\
PMM 351      & 08 36 24.2 & -54 01 06 &  90 &  6064 &  2.87 &  90   \\
PMM 3359     & 08 36 55.0 & -53 08 34 & 226 &  5460 &  2.81 &   9   \\
PMM 5376     & 08 37 02.3 & -52 46 59 &   0 &  3999 &       &  10   \\
PMM 4324     & 08 37 47.0 & -52 52 12 &  97 &  6225 &  3.02 &  41   \\
PMM 665      & 08 37 51.6 & -53 45 46 & 189 &  5506 &  2.76 &   8   \\
PMM 4336     & 08 37 55.6 & -52 57 11 &     &  4964 &       &   8   \\
PMM 4362     & 08 38 22.9 & -52 56 48 & 191 &  5797 &  3.02 &   9   \\
PMM 4413     & 08 38 55.7 & -52 57 52 & 163 &  5797 &  2.94 &   9   \\
PMM 686      & 08 39 22.6 & -53 55 06 & 215 &  4633 &  1.91 &  13   \\
PMM 4467     & 08 39 53.0 & -52 57 57 & 190 &  5140 &  2.41 &  13   \\
PMM 1083     & 08 40 06.2 & -53 38 07 & 162 &  5988 &  3.09 &  68   \\
PMM 8415     & 08 40 16.3 & -52 56 29 & 302 &  5059 &  2.56 &  21   \\
PMM 1759     & 08 40 18.3 & -53 30 29 &  55 &  4158 &  0.61 &   4   \\
PMM 1142     & 08 40 49.1 & -53 37 45 &     &  5567 &       &   8   \\
PMM 1174     & 08 41 22.7 & -53 38 09 &  79 &  6556 &  3.14 &  60   \\
PMM 1820     & 08 41 25.9 & -53 22 41 & 333 &  4524 &  1.99 &  86   \\
PMM 4636     & 08 41 57.8 & -52 52 14 & 100 &  4139 &  0.86 &   5   \\
PMM 3695     & 08 42 18.6 & -53 01 57 &   0 &  3661 &       &  90   \\
PMM 756      & 08 43 00.4 & -53 54 08 & 217 &  5383 &  2.72 &  16   \\
PMM 5811     & 08 43 17.9 & -52 36 11 & 107 &  6760 &  3.40 &  59   \\
PMM 2888     & 08 43 52.3 & -53 14 00 &     &  6483 &       &  66   \\
PMM 2012     & 08 43 59.0 & -53 33 44 & 251 &  5168 &  2.58 &  17   \\
PMM 4809     & 08 44 05.2 & -52 53 17 & 175 &  5759 &  2.94 &  18   \\
PMM 1373     & 08 44 10.2 & -53 43 34 & 156 &  4785 &  1.94 &   7   \\
PMM 5884     & 08 44 26.2 & -52 42 32 & 225 &  5383 &  2.73 &  14   \\
PMM 4902     & 08 45 26.9 & -52 52 02 & 204 &  4508 &  1.73 &   8   \\
PMM 6811     & 08 45 39.1 & -52 26 00 & 137 &  6330 &  3.26 &  90   \\
PMM 2182     & 08 45 48.0 & -53 25 51 & 187 &  6094 &  3.24 &  78   \\
CD-57 2315   & 08 50 08.1 & -57 45 59 & 308 &  4917 &  2.41 &  24   \\
TYC 8594-0058-1 & 09 02 03.9&-58 08 50& 300 &  5432 &  2.93 &  34   \\
CD-62 1197   & 09 13 30.3 & -62 59 09 & 280 &  5083 &  2.54 &  84   \\
TYC 7695-0335-1 & 09 28 54.1&-41 01 19& 300 &  4735 &  2.19 & 120   \\
HD 84075     & 09 36 17.8 & -78 20 42 & 170 &  6216 &  3.28 &  20     \\
TYC 9217-0641-1 & 09 42 47.4&-72 39 50& 240 &  5083 &  2.47 &  23   \\
CD-39 5883   & 09 47 19.9 & -40 03 10 & 260 &  5225 &  2.65 &  11  \\
HD  85151A   & 09 48 43.3 & -44 54 08 & 220 &  5607 &  2.93 &         \\
HD  85151B   & 09 48 43.5 & -44 54 09 & 250 &  5383 &  2.79 &         \\
CD-65 817    & 09 49 09.0 & -65 40 21 & 200 &  5988 &  3.19 &  19   \\
HD 309851    & 09 55 58.3 & -67 21 22 & 170 &  6026 &  3.14 &  19   \\
HD 310316    & 10 49 56.1 & -69 51 22 & 224 &  5645 &  2.97 &  16   \\
CP-69 1432   & 10 53 51.5 & -70 02 16 & 195 &  5912 &  3.12 &  55   \\
CD-74 673    & 12 20 34.4 & -75 39 29 & 230 &  4616 &  1.92 &  12     \\
CD-75 652    & 13 49 12.9 & -75 49 48 & 200 &  5721 &  2.98 &  20   \\
HD 129496    & 14 46 21.4 & -67 46 16 & 150 &  6330 &  3.30 &  88   \\
NY Aps       & 15 12 23.4 & -75 15 16 & 182 &  5304 &  2.55 &  11   \\
CD-52 9381   & 20 07 23.8 & -51 47 27 &  60 &  4168 &  0.67 &  42   \\
\hline
\multicolumn{7}{||c||} {Intruders}\\
\hline
TYC 6585-0334-1&08 56 26.3 & -22 41  40 &  10 & 4226 & -0.06      & 4     \\
\hline\hline	                                                    
\end{tabular}                                                       
\end{table*} 
\begin{table*}
\caption{The AB Doradus Association}
\begin{tabular}{||lrrrrrr||}
\hline\hline
Ident & $\alpha$(2000) & $\delta$(2000) & EW$_{\rm Li}$ & T$_{\rm eff}$ & A$_{\rm Li}$& $v\sin(i)$ \\
~ & ~& ~ & m{\AA} &  K  & ~& km~s$^{-1}$ \\
\hline
PW And       & 00 18 20.9 & +30 57 22 & 267 &  4701 &  2.10 &  22    \\
HD 4277      & 00 45 50.9 & +54 58 40 & 119 &  6521 &  3.32 &  24      \\
HD 6569      & 01 06 26.2 & -14 17 47 & 155 &  5036 &  2.21 &  10      \\
CD-12 243    & 01 20 32.3 & -11 28 04 & 160 &  5432 &  2.62 &   3    \\
CD-46 644    & 02 10 55.4 & -46 03 59 & 250 &  4629 &  1.90 &  36    \\
HD 13482     & 02 12 15.4 & +23 57 29 & 145 &  5597 &  2.65 &   6      \\
HIP 12635    & 02 42 21.0 & +38 37 21 & 146 &  4917 &  2.05 &   6      \\
HD 16760A    & 02 42 21.3 & +38 37 07 & 158 &  5829 &  2.96 &   3      \\
HD 17332B    & 02 47 27.2 & +19 22 21 & 170 &  5721 &  2.90 &   8      \\
HD 17332A    & 02 47 27.4 & +19 22 19 & 155 &  6064 &  3.13 &  13      \\
IS Eri       & 03 09 42.3 & -09 34 47 & 191 &  5383 &  2.66 &   7    \\
HIP 14807    & 03 11 12.3 & +22 25 23 &  34 &  4226 &  0.49 &          \\
HIP 14809    & 03 11 13.8 & +22 24 57 & 145 &  6102 &  3.13 &          \\
BD+09 412    & 03 12 34.3 & +09 44 57 & 150 &  4735 &  1.86 & 6         \\
V577 Per     & 03 33 13.5 & +46 15 27 & 200 &  5729 &  2.99 &   7      \\
HD 21845B    & 03 33 14.0 & +46 15 19 &  30 &  3831 & -0.28 &  20      \\
HIP 17695    & 03 47 23.3 & -01 58 20 &   0 &  3518 &       &  18      \\
HD 24681     & 03 55 20.4 & -01 43 45 & 230 &  5630 &  2.97 &  34      \\
HD 25457     & 04 02 36.7 & -00 16 08 & 100 &  6406 &  3.16 &  18      \\
HD 25953     & 04 06 41.5 & +01 41 02 & 120 &  6521 &  3.32 &  30      \\
TYC 0091-0082-1 & 04 37 51.5 & +05 03 08 & 220 &  5225 &  2.57 &  8        \\
TYC 5899-0026-1 & 04 52 24.4 & -16 49 22 &  20 &  3480 & -1.09 &     5     \\
CD-56 1032B  & 04 53 30.5 & -55 51 32 &   0 &  3388 &       &          \\
CD-56 1032A  & 04 53 31.2 & -55 51 37 &   0 &  3458 &       &          \\
HD 31652     & 04 57 22.3 & -09 08 00 & 230 &  5460 &  2.82 &   6       \\
CD-40 1701   & 05 02 30.4 & -39 59 13 & 120 &  4589 &  1.59 &   7    \\
HD 32981     & 05 06 27.7 & -15 49 30 & 140 &  5995 &  3.03 &     6     \\
HD 293857    & 05 11 09.7 & -04 10 54 & 260 &  5481 &  2.90 &  11      \\
HD 33999     & 05 12 35.8 & -34 28 48 &  90 &  6523 &  3.19 &   7    \\
HD 35650     & 05 24 30.2 & -38 58 11 &  15 &  4165 &  0.02 &   4    \\
AB DorB      & 05 28 44.4 & -65 26 47 &   0 &  3397 &       &          \\
AB DorA      & 05 28 44.8 & -65 26 56 & 267 &  5059 &  2.50 &  53      \\
UX Col       & 05 28 56.5 & -33 28 16 & 182 &  4491 &  1.65 &  42    \\
CD-34 2331   & 05 35 04.1 & -34 17 52 & 190 &  4589 &  1.80 &  25    \\
HIP 26369    & 05 36 55.1 & -47 57 48 &  70 &  4197 &  0.79 &  28    \\
UY Pic       & 05 36 56.9 & -47 57 53 & 285 &  5197 &  2.68 &   9      \\
WX Col       & 05 37 12.9 & -42 42 56 & 180 &  5759 &  2.96 &   6    \\
HIP 26401B   & 05 37 13.2 & -42 42 57 & 270 &  5012 &  2.45 &   5    \\
Par 2752     & 05 38 56.6 & -06 24 41 & 130 &  5607 &  2.68 &   6      \\
CP-19 878    & 05 39 23.2 & -19 33 29 & 290 &  5083 &  2.57 &  32      \\
TYC 7605-1429-1   & 05 41 14.3 & -41 17 59 & 250 &  4589 &  1.84 &  54    \\
CD-26 2425   & 05 44 13.4 & -26 06 15 & 320 &  4917 &  2.44 &          \\
TZ Col       & 05 52 16.0 & -28 39 25 & 180 &  5988 &  3.15 &  20      \\
TY Col       & 05 57 50.8 & -38 04 03 & 210 &  5481 &  2.80 &  55    \\
BD-13 1328   & 06 02 21.9 & -13 55 33 & 230 &  4589 &  1.89 &  6        \\
CD-34 2676   & 06 08 33.9 & -34 02 55 & 240 &  5383 &  2.77 &   4    \\
CD-35 2722   & 06 09 19.2 & -35 49 31 &  10 &  3727 & -0.97 &  13    \\
HD 45270     & 06 22 30.9 & -60 13 07 & 150 &  6102 &  3.14 &  18    \\
GSC 8894-0426 & 06 25 56.1 & -60 03 27 &   0 &  3492 &       &          \\
AK Pic       & 06 38 00.4 & -61 32 00 & 140 &  6178 &  3.16 &  18    \\
CD-61 1439   & 06 39 50.0 & -61 28 42 &  40 &  4158 &  0.46 &  12      \\
TYC 7627-2190-1   & 06 41 18.5 & -38 20 36 & 280 &  4917 &  2.37 &  37    \\
GSC 8544-1037 & 06 47 53.4 & -57 13 32 & 170 &  4589 &  1.75 &   7      \\
CD-57 1654   & 07 10 50.6 & -57 36 46 & 180 &  5931 &  3.10 &  29    \\
BD+20 1790   & 07 23 43.6 & +20 24 59 & 105 &  4394 &  1.27 &  13      \\
HD 59169     & 07 26 17.7 & -49 40 51 & 145 &  5607 &  2.73 &  11    \\
V372 Pup     & 07 28 51.4 & -30 14 49 &  10 &  4011 & -0.42 &  20    \\
CD-84 80     & 07 30 59.5 & -84 19 28 & 300 &  5196 &  2.70 &   3    \\
HD 64982     & 07 45 35.6 & -79 40 08 & 140 &  6330 &  3.27 &  14      \\
BD-07 2388   & 08 13 51.0 & -07 38 25 & 300 &  5390 &  2.89 & 130      \\
HD 82879     & 09 28 21.1 & -78 15 35 & 100 &  6623 &  3.29 & 140     \\
CD-45 5772   & 10 07 25.2 & -46 21 50 &  50 &  4589 &  1.18 &   6    \\
BD+01 2447   & 10 28 55.5 & +00 50 28 &   0 &  3570 &       &   0    \\
HD 99827     & 11 25 17.7 & -84 57 16 &  80 &  6711 &  3.24 &  45      \\
\hline\hline	                                                    
\end{tabular}                                                       
\end{table*} 
\addtocounter{table}{-1}
\begin{table*}
\caption{The AB Doradus Association - Continued}
\begin{tabular}{||lrrrrrr||}
\hline\hline
Ident & $\alpha$(2000) & $\delta$(2000) & EW$_{\rm Li}$ & T$_{\rm eff}$ & A$_{\rm Li}$& $v\sin(i)$ \\
~ & ~& ~ & m{\AA} &  K  & ~& km~s$^{-1}$ \\
\hline

PX Vir       & 13 03 49.7 & -05 09 43 & 142 &  5197 &  2.33 &   6      \\
HD 139751    & 15 40 28.4 & -18 41 46 & 110 &  4442 &  1.36 &   8      \\
HIP 81084    & 16 33 41.6 & -09 33 12 &   0 &  3879 &       &   7      \\
HD 152555    & 16 54 08.1 & -04 20 25 & 133 &  6102 &  3.08 &  16      \\
HD 317617    & 17 28 55.6 & -32 43 57 & 120 &  4735 &  1.76 &   4    \\
HD 159911    & 17 37 46.5 & -13 14 47 & 250 &  4589 &  1.93 & 140      \\
HD 160934    & 17 38 39.6 & +61 14 16 &  40 &  4226 &  0.57 &  17      \\
HD 176367    & 19 01 06.0 & -28 42 50 & 140 &  6254 &  3.22 &  17      \\
HD 178085    & 19 10 57.9 & -60 16 20 & 165 &  6102 &  3.19 &  24      \\
TYC 0486-4943-1    & 19 33 03.8 & +03 45 40 & 180 &  4735 &  1.95 &     11     \\
HD 189285    & 19 59 24.1 & -04 32 06 & 140 &  5630 &  2.73 &      9   \\
BD-03 4778   & 20 04 49.4 & -02 39 20 & 290 &  5083 &  2.57 &       8   \\
HD 199058    & 20 54 21.1 & +09 02 24 & 160 &  5894 &  3.01 &  14        \\
TYC 1090-0543-1 & 20 54 28.0 & +09 06 07 & 120 &  4589 &  1.59 &   18       \\
HD 201919    & 21 13 05.3 & -17 29 13 &  20 &  4268 &  0.32 &   8    \\
LQ Peg       & 21 31 01.7 & +23 20 07 & 215 &  4524 &  1.78 &  66    \\
HD 207278    & 21 48 48.5 & -39 29 09 & 190 &  5759 &  2.99 &  10    \\
HIP 107948   & 21 52 10.4 & +05 37 36 &   0 &  3486 &       &  80      \\
HIP 110526A  & 22 23 29.1 & +32 27 34 &   0 &  3492 &       &  16      \\
HIP 110526B  & 22 23 29.1 & +32 27 32 &   0 &  3480 &       &          \\
HD 217343    & 23 00 19.3 & -26 09 14 & 180 &  5912 &  3.09 &  12    \\
HD 217379    & 23 00 28.0 & -26 18 43 &  10 &  4257 & -0.01 &   6      \\
HIP 114066   & 23 06 04.8 & +63 55 34 &  30 &  3879 & -0.18 &   8      \\
HD 218860A  & 23 11 52.1 & -45 08 11 & 222 &  5607 &  2.94 &   7    \\
HD 218860B  & 23 11 53.6 & -45 08 00 &   0 &  3492 &       &          \\
HIP 115162   & 23 19 39.6 & +42 15 10 & 160 &  5607 &  2.78 &          \\
HD 222575    & 23 41 54.3 & -35 58 40 & 230 &  5567 &  2.92 &  31    \\
HD 224228   & 23 56 10.7 & -39 03 08 &  78 &  4826 &  1.66 &   3    \\ 
\hline
\multicolumn{7}{||c||} {Intruders}\\
\hline	 
CD-41 2076  &05 48 30.4 & -41  27 20 &  20 &   4361 & 0.45      &  10  \\
V402 Hya    &08 53 12.1 & -07  43 21 &   0 &   5111 &       &   240   \\
CD-37 6177  &09 56 58.4 & -38  33 14 &   0 &   4917 &       &      \\
HD 110810   &12 45 14.4 & -57  21 29 &   0 &   4964 &       &   4  \\
\hline \hline
\end{tabular}  
\end{table*}

\end{document}